\documentclass[12pt,a4paper,tightenlines,nofootinbib]{revtex4}
\usepackage{amssymb}
\usepackage{amsmath}
\usepackage{epsf}
\usepackage{bm}
\usepackage{graphicx}
\usepackage{pdfpages}
\usepackage{epstopdf}
\newcommand {\eps}{\epsilon}

\newcommand {\Up}{\Upsilon}
\newcommand {\si}{\sigma}
\newcommand {\ga}{\gamma}
\newcommand {\Ga}{\Gamma}
\newcommand {\de}{\delta}

\newcommand {\la}{\lambda}
\newcommand {\al}{\alpha}
\newcommand {\be}{\beta}
\newcommand {\pa}{\partial}

\newcommand {\ka}{\kappa}
\newcommand {\fr}{\frac}

\newcommand {\ca}{{\cal A}}

\newcommand {\ce}{{\cal E}}

\newcommand {\ch}{{\cal H}}

\newcommand {\cl}{{\cal L}}

\newcommand {\co}{{\cal O}}

\newcommand {\lan}{\langle}
\newcommand {\ran}{\rangle}

\newcommand {\bk}{\mathbf{k}}
\newcommand {\bx}{\mathbf{x}}

\newcommand {\lf}{\left}
\newcommand {\ri}{\right}
\newcommand {\beg}{\begin{equation}}
\newcommand {\en}{\end{equation}}
\newcommand {\bega}{\begin{eqnarray}}
\newcommand {\ena}{\end{eqnarray}}

\begin{document}

\author{Federico Agust\'{\i}n Membiela}

\email[]{membiela@mdp.edu.ar} \affiliation{Centro Brasileiro de
Pesquisas F\'{\i}sicas, Rua Xavier Sigaud, 150, Rio de Janeiro,
Brazil.} \affiliation{Departamento de F\'{\i}sica, Facultad de
Ciencias Exactas y Naturales,
  UNMdP, De\'{a}n Funes 3350, (7600) Mar del Plata, Argentina}

\title{Primordial magnetic fields from a non-singular bouncing cosmology}

\begin{abstract}

Although inflation is a natural candidate to generate the lengths
of coherence of magnetic fields needed to explain current
observations, it needs to break conformal invariance of
electromagnetism to obtain significant magnetic amplitudes. Of the
simplest realizations are the kinetically-coupled theories
$f^2(\phi)F_{\mu\nu}F^{\mu\nu}$ (or $IFF$ theories). However,
these are known to suffer from electric fields backreaction or the
strong coupling problem. In this work we shall confirm that such
class of theories are problematic to support magnetogenesis during
inflationary cosmology. On the contrary, we show that a bouncing
cosmology with a contracting phase dominated by an equation of
state with $p>-\rho/3$ can support magnetogenesis, evading the
{backreaction/strong-coupling problem}. Finally, we study safe
magnetogenesis in a particular bouncing model with an
ekpyrotic-like contracting phase. In this case we found that
$f^2(\phi)F^2$-instabilities might arise during the final
kinetic-driven expanding phase for steep ekpyrotic potentials.


\end{abstract}

\maketitle

\section{Introduction}

One of the open problems of modern cosmology is to explain the
origin of large scale magnetic fields in the structures of the
universe. Indeed, during the last two decades the refinement of
the experimental methods and the development of new ones allowed
the confirmation of galactic and extragalactic microgauss magnetic
fields. This was achieved through numerous observations
\cite{measuresB1}. Moreover, from recent observations of secondary
gamma rays produced by TeV-Blazars cosmic rays \cite{blazars},
there were derived lower and upper limits for B-fields in voids
regions: $1\times10^{-17}G<B_0<3\times10^{-14}G$.

The first theoretical explanation for galactic magnetic fields was
the dynamo mechanism \cite{dynamo}. The conventional dynamo is
supported by classical magnetohydrodynamics and it works on
galactic scales during the formation of the galaxy. However, this
idea has lost ground for achieving galactic B-fields just by
itself, and has been relegated to an amplification mechanism over
primordial magnetic fields. The dynamo acts on galactic scales so
in principle should only account for B-fields in galaxies. Thus,
for explaining the presence of fields in the intergalactic medium,
it needs additional mechanisms of ejection. Indeed, magnetic
fields have been reported in the large-scale structure
\cite{superclusters}. Another important argument against the
dynamo is the presence of microgauss fields in redshifted galaxies
or galaxies in formation \cite{redshift}. In these cases it is
difficult to explain how such strong B-fields could have been
generated in places where the dynamo had not enough time to act.
Finally, the dynamo mechanism needs the presence of initial
magnetic fields with minimum strengths and coherence lengths to be
efficient.

In turn, this led much of the theoretical efforts to develop
models of the early universe to account for the desired
magnetogenesis. Indeed, in this subject, one can find many
interesting large reviews with complete references therein
\cite{reviews}.

One of the many places where primordial magnetism is searched is
during inflationary cosmology. The main feature of inflation is
that it can naturally address the wide presence of magnetic
fields, specially on the large scale. However, it is known that
conformal symmetry of electromagnetism should be broken in order
to obtain appreciable magnetic fields to the end of inflation.
This was firstly done in \cite{Widrow88} by introducing new terms
that coupled the electromagnetic field to the curvature of the
universe. Another mechanism of breaking conformal invariance while
keeping the gauge symmetry was proposed in \cite{Ratra92} and
consisted in coupling the electromagnetic fields to a scalar field
through $e^{\la\phi}F^2$. This last scenario was inspired by
string theories of gravity \cite{Veneziano95}. In \cite{Fogli08} a
coupling $R^nF^2$ was studied during inflation.
Furthermore, in \cite{Martin07} a complete analysis was performed,
with models inspired by particle physics. Besides, at that time,
instabilities in the background dynamics risen by the backreaction
of the gauge field were already considered. However, until the
study \cite{Mukhanov09}, the strong coupling problem was ignored.
Here it was realized that inflationary magnetogenesis scenarios,
that evade the backreaction of electric fields, could not evade
very large values of the effective coupling 'constant' during its
initial stage.

However, inflation is not the only early cosmological theory that
generates the stretching of microphysical fluctuations to
super-Hubble scales. One possible scenario is a bouncing cosmology
\cite{Novello}. Yet, contracting cosmological phases are known to
suffer from BKL-chaos instability generated by anisotropies
\cite{BKL}. Indeed, their energy density grows as fast as
$a^{-6}$, like a stiff fluid (with $w=1$). In order to avoid such
instabilities, one needs in the Friedmann equations a stiffer
component (with $w>1$) for the background universe. In particular,
this is done by the ekpyrotic scenario \cite{ekpyrotic1}. There
are many cosmological models inspired in such scenario, like
cyclic cosmologies \cite{cyclic} and the "new ekpyrotic"
cosmologies \cite{Ovrut07} [see \cite{ekpyrotic2} for a recent
review]. Other realizations are \cite{many bounce}, where new
physics are required to provide the bounce, by violating the null
energy condition. Particularly, in new ekpyrotic scenario the job
is done by a ghost condensate phase \cite{Creminelli06} of the
scalar field, yielding a non-singular bounce.

A previous study concerning early magnetogenesis from bouncing
cosmologies is \cite{Salim0612}. However, they considered a
background were the effective equation of state is $w=1$, thus it
is not clear if the contracting phase is free of the anisotropy
instability. Additionally, electromagnetic instabilities or
backreaction were not studied. And finally, the coupling function
$I(\omega)=e^{-2\omega}$ remains always well below unity for their
parameters $\la\approx 0.07\sim0.1$, thus yielding a
strong-coupled regime.

The main objective of the present work is to identify the
condition which determine the presence of the strong
coupling/backreaction problem in a cosmological background. We
start by considering an early background cosmological phase
dominated by a constant equation of state $w$. Here we placed the
$f^2F^2$ (or $IFF$) electromagnetic theory at a perturbative
level. Initially, the coupling function is defined through a
power-law $f\propto a^{n}$, where we call '$n$' the coupling
parameter\footnote{In the work \cite{Martin07} notation $\al$ was
used for the coupling parameter.}. This parametrization is useful
since it leads to simple power-law spectrum for the
electromagnetic field. Using this, we find a generalized parameter
$\ga=2n/(1+3w)$, that characterizes the dynamics of the
electromagnetic modes and its spectrum.
Next we considered the backreaction of electromagnetic fields. As
already noted in previous works, we find that the branch safe from
backreaction of the electric fields corresponds to $\ga<0$. Here
B-fields lead the spectrum of electromagnetic fields and make
magnetogenesis efficient. However, the strong-coupled regime
belongs to values $n>0$. Thus, one confirms that during inflation
both problems cannot be simultaneously solved. Clearly, such a
statement is done for the present $f^2F^2$ universality, without
introducing any further assumptions (in particular in
\cite{Sloth13} they considered low-scale inflation). Nevertheless,
to simultaneously address both problems, one needs that $\ga<0$
and $n<0$. We find that this will only happen for cosmological
phases with $w>-1/3$.

This condition motivated the second part of our work. We searched
for magnetogenesis in an early cosmological phase with $w>-1/3$.
This phase should have the property of stretching microphysical
fluctuations to super-Hubble scales, similarly to inflation. Such
condition can only be achieved by contracting phases ($\dot a<0$
or $H<0$) given that their comoving Hubble sphere shrinks like
$(a|H|)^{-1}\propto a^{(1+3w)/2}$ (for $w>-1/3$).

We used the background cosmology developed by Yi-Fu Cai, et.al. in
\cite{Bran1206} as an example to study the production of magnetic
fields. This model addresses the almost scale invariant spectrum
of primordial inhomogeneities by using curvature perturbations
instead of entropy perturbations, as is done in the usual
ekpyrotic scenarios. This model is a realization of the
\emph{Matter-Bounce} scenarios \cite{Bran0904} that includes an
early contracting phase with an ekpyrotic-like equation of state
($w>1$). Indeed, this contracting phase is refereed as the
\emph{ekpyrotic phase}. Also, in \cite{Bran1301} it was studied,
at the linear level, that such a realization was stable against
chaos from anisotropy during the bounce. This model was further
extended to a two-field picture in \cite{Bran1305} . Moreover,
very recently this was confirmed at the nonperturbative level
\cite{Steinhardt1308}, but showing some fine-tunning problems.

We checked that this model is free from the backreaction/strong
coupling problem during the contracting phase. Indeed, the weak
coupled regime persist during the whole dynamics. However, we find
that instabilities might arise during the final fast-roll
expanding phase. This backreaction could be severe if the
fast-roll lasts for too long when the ekpyrotic potential is very
steep and asymmetric. Yet, for weaker ekpyrotic phases stability
is recovered.


The paper is organized as follows. In Section\ref{Sec2} we study
the evolution of electromagnetic fields kinetically-coupled to a
scalar field by $f^2(\phi)F^2$. We assume the scalar field
dominates the Einstein equations, thus the electromagnetic field
is effectively coupled to the background dynamics. We study the
conditions on the parameters where the backreaction/strong
coupling problem appears. In Section\ref{Bfields} we apply these
results to a specific example that deploys a bounce. It starts
from an early ekpyrotic contracting phase and then bounces to a
fast-roll expanding period. We set initial conditions to the
electromagnetic field during the contracting phase. Then, we study
the evolution of these modes through the different phases. In
doing so, we introduce an specific model for the coupling of the
$e^{\la\phi}$ type. This is consistent with the background
dynamics and the power-law behavior. Next we kept the track of the
levels of backreaction during the contracting an the expanding
phase. Finally in Section\ref{Conclu} we give our conclusions. In
this article we use Natural units: $c=\hbar=k_B=1$.

\section{$f(\phi)^2F^2$ Electromagnetic fields in a cosmological
background}\label{Sec2}

We adopt a simple model where the electromagnetic field is coupled
to a scalar field that dominates the energy density of the
universe. This coupling is referred to as a $f(\phi)^2F^2$ or
$IFF$-class models. Indeed, as $\phi$ varies with time this
implies that, effectively, we are introducing a time-dependent
coupling 'constant' $e_\mathrm{eff}$, given by the inverse of the
coupling function, $e_\mathrm{eff}=e f^{-1}$. This kind of
scenarios had been extensively studied during an inflationary
phase to account for the generation of primordial magnetic fields.
However, they have proven to suffer two kinds of troubles:
\textbf{the strong coupling problem} and \textbf{the backreaction
problem} \cite{Mukhanov09}. Even worse, where one can be solved
the other emerges.

Other realizations were kinetic couplings have been recently used
are for: anisotropic-inflation or gauge-inflation
\cite{Sodareport}, preheating of the inflaton with a U(1) gauge
field \cite{Caldwell13} and non-Gaussianity features
\cite{Bnongaussian}.

We shall start by studying how the fluctuation modes of the gauge
field evolve in a cosmological background with a constant equation
of state $p=w\rho$. In doing so, we shall find a simple relation
between $w$, the spectrum of magnetic fields and the
backreaction/strong coupling problem.



We will adapt part of the derivations used in \cite{Martin07} for
inflation, and refer the reader there for detailed derivations.

\subsection{Electromagnetic perturbations coupled to the background dynamics}

Consider the action of a gauge $U(1)$ field (which we shall
identify with 'the electromagnetic field') coupled to a scalar
field $\phi$, as
    \beg\label{Sem}
        S_\mathrm{em}=-\fr{1}{4}\int
        d^4x\sqrt{-g}f^2(\phi)F_{\mu\nu}F^{\mu\nu},
    \en
with the electromagnetic tensor $F_{\mu\nu}=\pa_\mu A_\nu-\pa_\nu
A_\mu$. Here the vector potential has dimensions of mass
$[A_\mu]=M$. The coupling function $f(\phi)$ is in this sense
dimensionless. The equations of motion for $A_\mu$ are
 \beg
    \pa_\mu[\sqrt{-g}f^2(\phi)F^{\mu\nu}]=0.
 \en
We assume that the gauge fields are small enough so that the
background dynamics are entirely determined by the homogeneous
scalar field, thus we consider an homogeneous and isotropic FRWL
space-time,
 \beg
     ds^2=g_{\mu\nu}dx^\mu
     dx^\nu=-dt^2+a^2(t)d\bx^2=a^2(\tau)(-d\tau^2+d\bx^2).
 \en
In this the last expression we used conformal time $\tau=\int
a^{-1}dt $. With the coulomb gauge $A_0=\pa_i A_i=0$, we
completely fix the gauge freedom. The vector field is promoted
through canonical quantization to an operator with Fourier
decomposition
 \beg
    A_i(\tau,\bx)=\int\fr{d^3k}{(2\pi)^{3/2}}\sum_{\la=1}^2
    \left[\eps_{i\la}(\bk)b_\la(\bk)A(\tau,k)e^{i\bk\cdot\bx}+h.c.\right].
 \en
Using the mode definition $u_k(\tau,k)=a(\tau)A(\tau,k)$,
\footnote{The polarization vectors $\eps_{i\la}$ are proportional
to $a$} we obtain a damped harmonic oscillator equation
 \beg\label{Ec uk}
    u_k''+2\fr{f'}{f}u'_k+k^2u_k=0,\hspace{1cm} \ddot u_k+\left(H+2\fr{\dot f}{f}\right)\dot
u_k+\fr{k^2}{a^2}u_k=0,
 \en
with time-dependent damping. In particular, for the value
$f=1=cte$ one recovers the conformal Maxwell theory. This equation
may also be written as
 \beg\label{Ak}
    (f u_k)''+\left(k^2-\fr{f''}{f}\right)(f u_k)=0.
 \en
Thus, sometimes is useful to consider the redefined modes
$\ca_k\equiv f u_k$. With the aid of which we can identify short
wavelength modes as those for which $k^2\gg f''/f$. Indeed, these
solutions behave like plane waves in Minkowski space, addressing
the vacuum initial conditions
 \beg\label{UV}
    u_k(\tau) \rightarrow\fr{1}{f \sqrt{2k}}e^{-ik\tau},\hspace{1cm}
    \ca_k(\tau) \rightarrow\fr{1}{\sqrt{2k}}e^{-ik\tau}.
 \en
On the other side, when this modes overpass the Hubble scale they
eventually get into the long wavelength regime $k^2\ll f''/f$,
where the fields stop oscillating and have the solution
 \beg
    u_k(\tau)\rightarrow
    c_1(k)+c_2(k)\int\fr{d\tau}{f^2},\hspace{1cm}\ca_k(\tau)\rightarrow
    c_1(k)f+c_2(k)f\int\fr{d\tau}{f^2}.
 \en
For $u_k$, the solution consists in a constant mode and a
time-dependent mode. We shall assume a power-law behavior,
 \beg\label{F}
    f(a)=f_\star\left(\fr{a}{a_\star}\right)^n,
 \en
where '$n$' is the \emph{coupling parameter}. Such a choice is
motivated by simplicity, as we shall see, this form gives a simple
power-law for the electromagnetic field spectrum. In this sense,
any simple model that has a simple power-law electromagnetic field
spectrum will be consistent with this parametrization. Moreover,
this choice also considers exponential forms as originally
introduced in \cite{Ratra92}. Defining the equation of state of
the universe as $p=w\rho$, we obtain that
 \bega
 \left\{%
\begin{array}{ll}
    a\propto t^\fr{2}{3(w+1)}, & \hbox{$w\neq-1$;} \\
    a\propto e^{H_It}, & \hbox{$w=-1$.} \\
\end{array}%
\right.
  \ena
This is sufficient to yield the solutions
 \bega\label{Sol1}
     u_k(\tau)&\simeq&
     c_1(k)+c_2(k)\left(\fr{a}{a_\star}\right)^r,\hspace{1cm}r=-2n+\fr{1}{2}(1+3w)\\
 \nonumber    \ca_k(\tau)&\simeq& c_1(k)\left(\fr{a}{a_\star}\right)^n+c_2(k)\left(\fr{a}{a_\star}\right)^{r+n},
 \ena
the first has a constant mode $c_1(k)$ and a time-dependent mode
with pivot amplitude $c_2(k)$.

During de Sitter inflation we have $w=-1$ that yields on the
solution for the vector modes (\ref{Sol1}) an exponent given by
 \beg
    r^{(in)}=-2n-1.
 \en
If now we turn to an ekpyrotic phase with a equation of state
$w=-1+2/(3q)$, we obtain an exponent
 \beg\label{alphaEkpy}
   r^{(ek)}=-2n-1+\fr{1}{q}.
 \en
This last expression can also be identified with slow-roll
inflation. Indeed, the slow-roll parameter is $\eps=1/q=3(1+w)/2$
for $w\approx-1$ or $q\gg1$. Conversely, this is the same
fast-roll parameter for ekpyrosis as $w\gg1$ and then $q\ll1$.
This is a manifestation of the duality (at linear level) between
inflation and ekpyrosis \cite{Turok0403}.

We start by noticing that $f'/f=n\ch$, where $\ch=a'/a=\dot a$ is
the conformal Hubble parameter. Furthermore, the scale factor is
related to conformal time $\tau=\int a^{-1}dt $, in terms of the
equation of state through
 \beg\label{a de tau}
    \fr{a}{a_\star}=\left(\fr{\tau}{\tau_\star}\right)^{\fr{2}{1+3w}},
 \en
and the Hubble parameter
 \beg\label{aH}
   \ch=aH=\fr{2}{1+3w}\tau^{-1}.
 \en
It is convenient to introduce a new parameter that depends both on
$n$ and $w$,
 \beg\label{gamma}
  \ga=\fr{2n}{1+3w},
 \en
with the aid of which the mode equation gets simplified
  \beg\label{ec uk2}
    {u_k}''+\fr{2\ga}{\tau}{u_k}'+k^2u_k=0,\hspace{1cm}
    \ca_k''+\left[k^2-\fr{\ga(\ga-1)}{\tau^2}\right]\ca_k=0.
 \en
One can either solve any of these last equations, and the exact
solution is given in terms of Bessel functions
 \beg\label{sol u}
    \ca_k(x)=x^{\fr{1}{2}}\left[ d_1(k) \ch^{(1)}_{\fr{1}{2}-\ga}(x) +
            d_2(k)\ch^{(2)}_{\fr{1}{2}-\ga}(x)\right],
 \en
with $x=-k\tau$. We have used Hankel functions for convenience, in
this sense the initial vacuum conditions [see eq.(\ref{UV})] are
trivially achieved for the second Hankel function
$\ch^{(2)}_{\ga-\fr{1}{2}}(x\rightarrow\infty)\rightarrow
x^{-1/2}e^{-ix}$. Thus, we obtain
 \bega
    d_1(k)&=&0,\\
    d_2(k)&=&\sqrt{\fr{\pi}{4k}}e^{i\fr{\pi}{2}(\ga-1)}.
 \ena
On the other side, the long wavelength regime for Hankel functions
may be computed from the asymptotic form of First Kind Bessel
functions $J_\mu(x)\propto x^\mu$. Moreover, we use that
$\ch^{(2)}_{\mu}(x)=i\csc(\al\pi)[J_{-\mu}(x)-e^{i\pi\mu}J_\mu(x)]$.
The expression for $u_k$ is then,
\beg\label{Sol2}
    u_k(x)=\fr{\ca_k}{f}=\fr{b_{(\ga)}}{f\sqrt{k}}x^{\ga}+\fr{b_{(1-\ga)}}{f\sqrt{k}}x^{1-\ga},
 \en
with the function
 \beg\label{Cte b}
    b_{(\ga)}=\fr{\pi^{\fr{1}{2}} e^{i\fr{\pi\ga}{2}}
    }{2^{\fr{1}{2}+\ga}\cos{(\ga\pi)}\Ga\lf(\fr{1}{2}+\ga\ri)}.
 \en
With this considerations one can determine exactly the constants
of the solution (\ref{Sol1}), that they show to be
 \bega\label{Cte}
    c_1(k)&=&b_{(\ga)}
    \left(\fr{2}{1+3w}\right)^\ga\fr{1}{f_\star{k}^{1/2}}\left(\fr{k}{a_\star
    H_\star}\right)^\ga,\\
  \nonumber  c_2(k)&=&b_{(1-\ga)}\left(\fr{2}{1+3w}\right)^{1-\ga}\fr{1}{f_\star{k}^{1/2}}\left(\fr{k}{a_\star
    H_\star}\right)^{1-\ga},
 \ena
where we have used (\ref{aH}) to change from the conformal pivot
time $\tau_\star$ to $a_\star H_\star$. Cleared up the constants,
we can apply a specific background model that fixes the value of
$w$, and by using (\ref{a de tau}), (\ref{aH}) and (\ref{gamma})
we are only left to determine the coupling function parameter $n$.
We shall either use $b_{(\ga)}, b_{(1-\ga)}$ or $c_1(k),c_2(k)$ in
convenience to alleviate the algebraic manipulations and notation.
Another important quantity is the time derivative of the $u_k$
modes. Not only it defines the electric energy density, but the
matching conditions as well. However, care should be taken because
the $c_1$-term is constant and so, when deriving, it would just
remain a contribution from the $c_2$-term. Yet, one has to
consider the next order in the power series of the Bessel
$J_\mu(x)$, that goes like $\sim x^{2+\mu}$. Thus, we arrive at
the expression,
 \beg\label{Sol dot uk}
    \dot
    u_k=\lf(\fr{1+3w}{2}\ri)H\lf[\fr{e_{(-\ga)}}{f\sqrt{k}}x^{\ga+2}+
    \fr{e_{(1+\ga)}}{f\sqrt{k}}x^{1-\ga}\ri],
 \en
where it has been used that $\dot x/x=(1+3w)H/2$ and the function
 \beg\label{Cte e}
    e_{(\ga)}=-\fr{b_{(-\ga)}}{1-2\ga}.
 \en

\subsection{Electromagnetic energy density}

For the determination of the magnetic fields amplitude and the
effects of backreaction that may occur, we shall compute the total
electromagnetic energy density stored at a certain scale $L=2\pi
k^{-1}$. The energy density is defined through the vacuum
expectation value of the time-time component of the stress tensor
as $\rho_\mathrm{em}\equiv-\lan0|{T}^0_0|0\ran$. The stress tensor
of the fields is determined by
 \beg
    T_{\mu\nu}\equiv -\fr{2}{\sqrt{-g}}\fr{\de S_\mathrm{em}}{\de
    g^{\mu\nu}}=-f^2(\phi)\left({F_\mu}^\al
    F_{\al\nu}+\fr{1}{4}g_{\mu\nu}F_{\al\be}F^{\al\be}\right).
 \en
We could further identify an electric and magnetic energy
densities. But first we need to define electric and magnetic
fields for relativistic observers with velocity $u^\al$,
 \beg
    E_\mu=u^\nu
    F_{\mu\nu},\hspace{1cm}B_\mu=\fr{1}{2}\eta_{\mu\nu\al\be}F^{\nu\al}u^\be,
 \en
where $\eta_{\mu\nu\al\be}=\sqrt{-g}\varepsilon_{\mu\nu\al\be}$ is
the space volume totally antisymmetric tensor related to the
Levi-Civita symbol in 4D. Therefore, for a comoving observer
$u^\al=(1,\mathbf{0})$, in cosmic time coordinates, one obtains an
electric field $E_i=-\dot{A}_i$ and a magnetic field
$B_i=a^{-1}\varepsilon_{ijk}\pa_j A_k$; both are written in the
usual 3D Euclidean notation (where the metric tensor is the
Kronecker Delta function $\de_{ij}$).

With the above definitions it is easy to see that
 \bega\label{T00}
     T_{0}^0=-\fr{f^2}{2a^2}\left(|\vec{E}|^2+|\vec{B}|^2\right).
 \ena
where $|\vec{E}|^2=\sum_{i=1}^3 E_i^2=a^2E_\mu E^\mu$ and
$|\vec{B}|^2=\sum_{i=1}^3 B_i^2=a^2B_\mu B^\mu$. This last
expression is written in cosmic time coordinates. One can pass to
conformal coordinates, but the electric field changes by a factor
$a$, then $\vec\ce=-\pa_\tau \vec A=a\vec E$.
 The energy density of magnetic fields stored at a given scale
 $L=2\pi/k$ is
 \beg\label{rhoB}
    \fr{d\rho_B}{d\ln k}=\fr{f^2}{2\pi^2}\fr{k^5}{a^4}|u_k|^2.
 \en
While the energy density of electric fields at a given scale is
 \beg\label{rhoE}
   \fr{d\rho_E}{d\ln k}=\fr{f^2}{2\pi^2}\fr{k^3}{a^2}|\dot u_k|^2.
 \en
Now we need to calculate the quadratic quantities $|u^{(ek)}_k|^2$
and $|\dot u^{(ek)}_k|^2$, so as to determine the magnetic
 and electric densities energies per unit
$k$. A direct calculation using the solution (\ref{Sol1}) or
(\ref{Sol2}) yields,
 \bega
\nonumber   |u_k|^2 &=& | c_1(k)|^2+| c_2(k)|^2 \left(
\fr{a}{a_{\star}}
\right)^{2r}+\lf[c_1(k)c_2^\ast(k)+c_1^\ast(k)c_2(k)\ri]\lf(\fr{a}{a_\star}\ri)^r\\
   &=&\fr{1}{f^2 k}\lf[|b_{(\ga)}|^2 x^{2\ga}+|b_{(1-\ga)}|^2
x^{2-2\ga}+(b_{(\ga)}b_{(1-\ga)}^\ast+b_{(\ga)}^\ast
b_{(1-\ga)})x\ri].
 \ena
The magnetic energy density on super Hubble scales ($x\ll1$) is
well approximated by
 \bega\label{drhoB12}
    \fr{d\rho_B}{d\ln
    k}=\fr{H^4}{2\pi^2}\lf(\fr{1+3w}{2}\ri)^4\left\{%
\begin{array}{ll}
    |b_{(\ga)}|^2
    (-k\tau)^{4+2\ga}, & \ga<\fr{1}{2} \\
    |b_{(1-\ga)}|^2 (-k\tau)^{6-2\ga}, & \ga>\fr{1}{2} \\
\end{array}%
 \ri.\ena
It is worth noticing that for a given spectral index there are two
possible values of $\gamma$. For example, if we seek for scale
invariant magnetic fields, then this is achieved for $\ga=3$, that
implies $n_B^{(2)}=6-2\ga=0$. But, also for $\ga=-2$, which yields
$n_B^{(1)}=4+2\ga=0$. Yet, one of this values may belong to the
strong-coupled regime and the other to the weak-coupled regime. In
the next we shall compare this possibilities between inflation and
ekpyrosis.

But first, it remains the determination of the electric energy
density per unit scale, that is defined through $\dot u_k$. Using
the solution (\ref{Sol dot uk}) in eq. (\ref{rhoE}) we get
 \beg\label{drhoE}
  \fr{d\rho_E}{d\ln
  k}=\fr{H^4}{2\pi^2}\lf(\fr{1+3w}{2}\ri)^4\left\{%
\begin{array}{ll}
    |e_{(-\ga)}|^2
    (-k\tau)^{6+2\ga}, & \ga<-\fr{1}{2} \\
    |e_{(1+\ga)}|^2 (-k\tau)^{4-2\ga}, & \ga>-\fr{1}{2} \\
\end{array}%
 \ri.
 \en
Similarly as the magnetic case, there are to possible values of
$\ga$ that yield the same spectrum.

\begin{figure}[h]
\centering
\includegraphics[scale=.65]{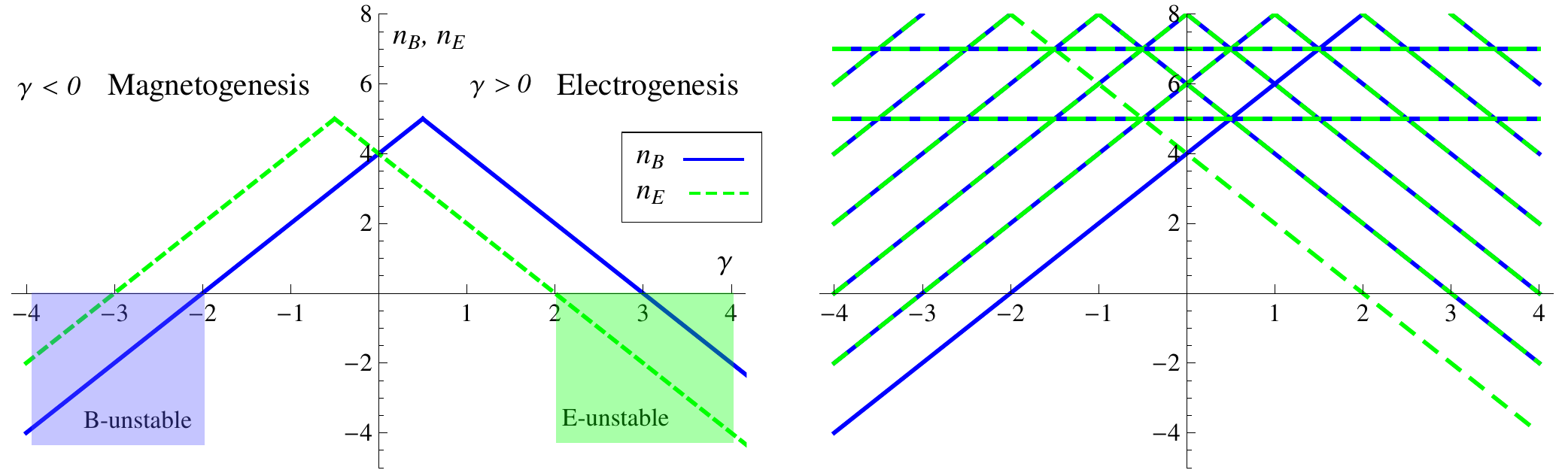}
\caption[Fig]{Left panel:  the leading spectral contributions from
the electric fields (green dashed line) and the magnetic fields
(blue line) versus $\ga$. Right panel: the overlap of sub-leading
$E$ and $B$-spectral indexes.} \label{index}
\end{figure}

One important observation about the last expressions
(\ref{drhoB12}) and (\ref{drhoE}) is that for a given value of
$\ga$ one of the terms will lead the spectrum, whereas the other
in general will not be the immediate in sub-leading order (unless
$|\ga|<1/2$). For example, if we take the value $\ga=-1$ the
leading B-spectrum is $n_B^{(1)}=2$. On the other side, the other
term gives a value $n_B^{(2)}=8$ that it is not the next to
leading order contribution. Indeed, there are many terms before
it, some of which are obtained by going to orders $k^2$ and $k^4$
starting from $n_B^{(1)}$ and others by considering the cross
terms between $c_1(k)$ and $c_2(k)$. This situation is detailed in
the right panel of Fig.(\ref{index}). Here we plotted the first
contributions in the spectrum from the electric and the magnetic
fields. Notice how the subleading contributions of electric and
magnetic fields will overlap. Additionally, the horizontal lines
corresponds to the cross terms contribution that start from the
lower B and E-spectrum $n_{BE}^{\mathrm{(cross)}}=5$, to higher.
Beside, on the left panel, we only kept the leading contribution
of them. The leading spectrum of each is defined in two 'branches'
of $\ga$, this was previously obtained in \cite{Martin07}.

Finally, a crucial observation is that for $\ga>0$ the E-spectrum
leads with respect to the B-spectrum. Then, in this region one
obtains \emph{electrogenesis}, since most of the energy is given
to electric fields in larger scales. For $\ga\geq1/2$ the
B-spectrum is blue tilted respect to E as $n_B=n_E+2$. When one
searches at inflation for the generation of magnetic fields in
large scales, this region, $\ga>0$, is associated with strong
backreaction \cite{Mukhanov09}. In particular, interesting
B-spectra are those close to scale invariance, $n_B\sim0$. In this
case, one gets a red-tilted spectrum for electric fields
$n_E\sim-2$. In the middle, the value $\ga=0$ yields blue magnetic
and electric fields with the same index: $n_B=n_E=4$. Of course,
this corresponds to the case $f=1=cte$, yielding the usual
electromagnetic fields energy density that decays like $a^{-4}$.
On the other side, the negative semiplane $\ga<0$ has a red tilted
B-spectrum with respect to E-fields. Moreover, for $\ga<-1/2$ the
red tilt is constant $n_E=n_B+2$. Therefore, this region should be
identified to produce \emph{magnetogenesis}. The value $\ga=-2$
yields scale invariant B-fields and a blue electric field with
$n_E=2$.

\subsection{Unstable growth of electromagnetic perturbations}

When writing the Friedmann equations one should be aware that
problems can arise if the perturbations of the vector field can
grow to energy levels comparable to the background. Indeed,
electromagnetic fields could generate serious deviations of
isotropy, given that their anisotropy is of the same order of
isotropy. In this sense, for the model to be successful it should
control its backreaction in the cosmological scales.

To consider backreaction we have to sum the collective
contribution of all the modes that reach the long wavelength
regime and have common initial sub-horizon conditions. Whereas
short wavelengths modes that remain below $(aH)^{-1}$ are
renormalized at the leading order.
 \bega
   \rho_E+ \rho_B=\int_{a_iH_i}^{aH}\fr{dk}{k} \left[\fr{d\rho_E}{d\ln k}+\fr{d\rho_B}{d\ln k}\right]
 \ena
Next we compare the energy densities $\rho_B$ and $\rho_E$ with
the background energy density $\rho_T=\fr{3}{8\pi}m_{pl}^2H^2$,
obtaining the ratios
 \bega\label{OmegaB1}
    \Omega_B^{(1)}\equiv\fr{\rho_B^{(1)}}{\rho_T}=\fr{H^2}{m_{pl}^2}\fr{4|b(\ga)|^2}{\pi}
    \left|\fr{2}{1+3w}\right|^{2\ga}
                    \times\\
                    \left\{%
\begin{array}{ll}
   \nonumber \fr{1}{4+2\ga}\left[1-e^{-(4+2\ga)N(t)}\right], & \hbox{$\ga\neq-2$;} \\
    N(t), & \hbox{$\ga=-2$.} \\
\end{array}%
\right.
  \ena
where we have used
$N(t)\equiv\ln\left(\fr{a(t)H(t)}{a_iH_i}\right)$ that accounts
for the number of e-folds\footnote{This quantity measures the
number of e-folds of (aH), in particular for inflation ($H\simeq
cte$) it will be related to e-folds of expansion.} measured from a
particular time $t_i$, when we have a variable $H$. In a similar
way we find that
 \bega
   \Omega_B^{(2)}\equiv\fr{\rho_B^{(2)}}{\rho_T}=\fr{H^2}{m_{pl}^2}\fr{4|b(1-\ga)|^2}
   {\pi}
   \left|\fr{2}{1+3w}\right|^{2-2\ga}
                    \times\\
                    \left\{%
\begin{array}{ll}
   \nonumber \fr{1}{6-2\ga}\left[1-e^{-(6-2\ga)N(t)}\right], & \hbox{$\ga\neq3$;} \\
    N(t), & \hbox{$\ga=3$.} \\
\end{array}%
\right.
 \ena
The label $(1)$ or $(2)$ refers to the $n_B^{(1)}$ and $n_B^{(2)}$
respectively. We also compute the relative electric energy density
 \bega
   \Omega_E^{(1)}\equiv \fr{\rho_{E}}{\rho_T}=\fr{H^2}{m_{pl}^2}\fr{4|e(-\ga)|^2}{\pi}
    \left|\fr{2}{1+3w}\right|^{2\ga}
                    \times\\
                                    \left\{%
\begin{array}{ll}
   \nonumber \fr{1}{6+2\ga}\left[1-e^{-(6+2\ga)N(t)}\right], & \hbox{$\ga\neq-3$;} \\
    N(t), & \hbox{$\ga=-3$.} \\
\end{array}%
\right.\\
   \Omega_E^{(2)}\equiv \fr{\rho_{E}}{\rho_T}=\fr{H^2}{m_{pl}^2}\fr{4|e(1+\ga)|^2}{\pi}
    \left|\fr{2}{1+3w}\right|^{2-2\ga}
                    \times\\
                                    \left\{%
\begin{array}{ll}
   \nonumber \fr{1}{4-2\ga}\left[1-e^{-(4-2\ga)N(t)}\right], & \hbox{$\ga\neq2$;} \\
    N(t), & \hbox{$\ga=2$.} \\
\end{array}%
\right.
 \ena
For the model to be free of backreaction of the electromagnetic
field, one needs that the last ratios remain well below unity
during the whole dynamics, $\Omega_B+\Omega_E\ll1$.

From the last ratios we can identify that there will be two types
of instabilities. One produced by electric fields ($\ga>2$) and
the other by magnetic fields ($\ga<-2$) see left panel of
Fig.\ref{index}). This happens essentially when the leading
spectral index (E or B) becomes negative or red-tilted. In that
case the exponentials factors grow with $N$.

\subsection{The inflationary phase: $w\simeq-1$}

Lets adopt the last derivations for the simplest inflationary
case: de Sitter inflation. Indeed, the generalization to a
slow-rolling model should be immediate. Here, we have $w=-1$ and
the Hubble parameter is constant $H=H_I$. In this case we obtain
from the parameter (\ref{gamma}) the value
 \beg
    \ga^{(\mathrm{in})}=-n
 \en
Thus, the solution [see eq.(\ref{Sol1})] for the modes is
$u_k^{(\mathrm{in})}=c_1^{(\mathrm{in})}(k)+c_2^{(\mathrm{in})}(k)\left(\fr{a}{a_f}\right)^{-2n-1}$
with the constants (\ref{Cte}) for values $\ga=-n$ and $w=-1$.
Furthermore, the spectra of electric and magnetic fields are
defined by the spectral indexes
 \beg
    n_B=\left\{%
\begin{array}{ll}
     n_B^{(1)}=4-2n, & n>-\fr{1}{2} \\
     n_B^{(2)}=6+2n, & n<-\fr{1}{2} \\
\end{array}%
\right.
   ,\hspace{1cm}n_E=
   \left\{%
\begin{array}{ll}
    n_E^{(1)}=6-2n, & n>\fr{1}{2} \\
    n_E^{(2)}=4+2n, & n<\fr{1}{2} \\
\end{array}%
 \right.\en
For the generation of large scale magnetic fields one pursues
almost scale invariant magnetic fields, more efficient to
concentrate energy on larger scales. Nevertheless, red-tilted
spectra, as we have seen, are likely to develop backreaction.

One obtains scale-invariant magnetic fields, $n_B=0$, for $\ga=3$
and $\ga=-2$. These corresponds with the values $n=-3$ and $n=2$,
respectively. Yet, the value $n=-3$ leaves us with a red-tilted
electric field $n_E=-2$, while for $n=2$, one gets a blue-tilt
$n_E=2$. This, in turn, means that the values close to $n=-3$ look
more problematic and inefficient, since most of the energy that
the mechanism carries to large scales goes to E-fields rather than
to B-fields. Indeed, trying to keep controlled the levels of
backreaction left us with very little magnetic fields
($10^{-32}$G) to the end of inflation ($N=75$), insufficient to
account for minimum levels needed by dynamo mechanisms
\cite{Mukhanov09}.

On the contrary, the value $n=2$ is safe from this drawback.
However, values $n>0$ belong to the strong-coupled regime. In
particular for $n=2$ and $N=75$, and given that to the end of
inflation $f_{\emph{end}}\sim1$, the inverse of the coupling
function is of order $e^{150}\sim10^{64}$ at the beginning of the
inflationary period, and the theory becomes non-perturbative and
all our analysis breaks down.

\subsection{The backreaction/strong-coupling problem: expansion or
contraction?}


Lets check expression (\ref{gamma}), this parameter $\ga$ defines
the spectrum of the electromagnetic field in a background with an
equation of state $w$ and with a coupling parameter $n$. We can
write $n/\ga=(1+3w)/2$. This expression is also present in the
relation for the comoving Hubble radius
 \beg
    (aH)^{-1}\propto a^{\fr{1+3w}{2}}
 \en
Thus, if the universe expands ($\dot a>0$) we have three cases.
When $w>-1/3$ the \emph{strong energy condition} (SEC) is achieved
and then the universe decelerates ($\ddot a<0$) and the Hubble
radius grows. The Big-Bang epochs dominated by ordinary matter
($w\gtrsim0$) belongs to this case. For $w<-1/3$ the SEC is
violated and the universe has an accelerated expansion with a
shrinking Hubble sphere. Slow-roll inflation for which
$w\simeq-1$, belongs to these class. Furthermore, the value
$w=-1/3$ yields a constant Hubble sphere.
 For a contracting universe ($\dot a<0$) the situation inverts.
When $w<-1/3$, the universe decelerates its contraction rate and
the comoving Hubble length grows with time. While, for $w>-1/3$,
the universe speeds up its contraction making the Hubble sphere
shrink. Finally, when $w=-1/3$, the Hubble sphere remains
constant.

Returning to the parameter $\ga$, we see that (independently if
the universe is expanding or contracting), when $w<-1/3$ the signs
of $n$ and $\ga$ will be inverted. Then, if we want to remain with
a dominant B-spectrum, red-tilted with respect to the E-spectrum,
we require that $\ga<0$. This automatically leaves us with a value
$n>0$, say, the strong-coupled regime [see Fig.(2)]. We find that
$(f^2F^2)-$class models in any expanding accelerated background
($w<-1/3$) cannot simultaneously avoid the backreaction and
strong-coupling problems.

\begin{figure}[h]
\centering
\includegraphics[scale=.5]{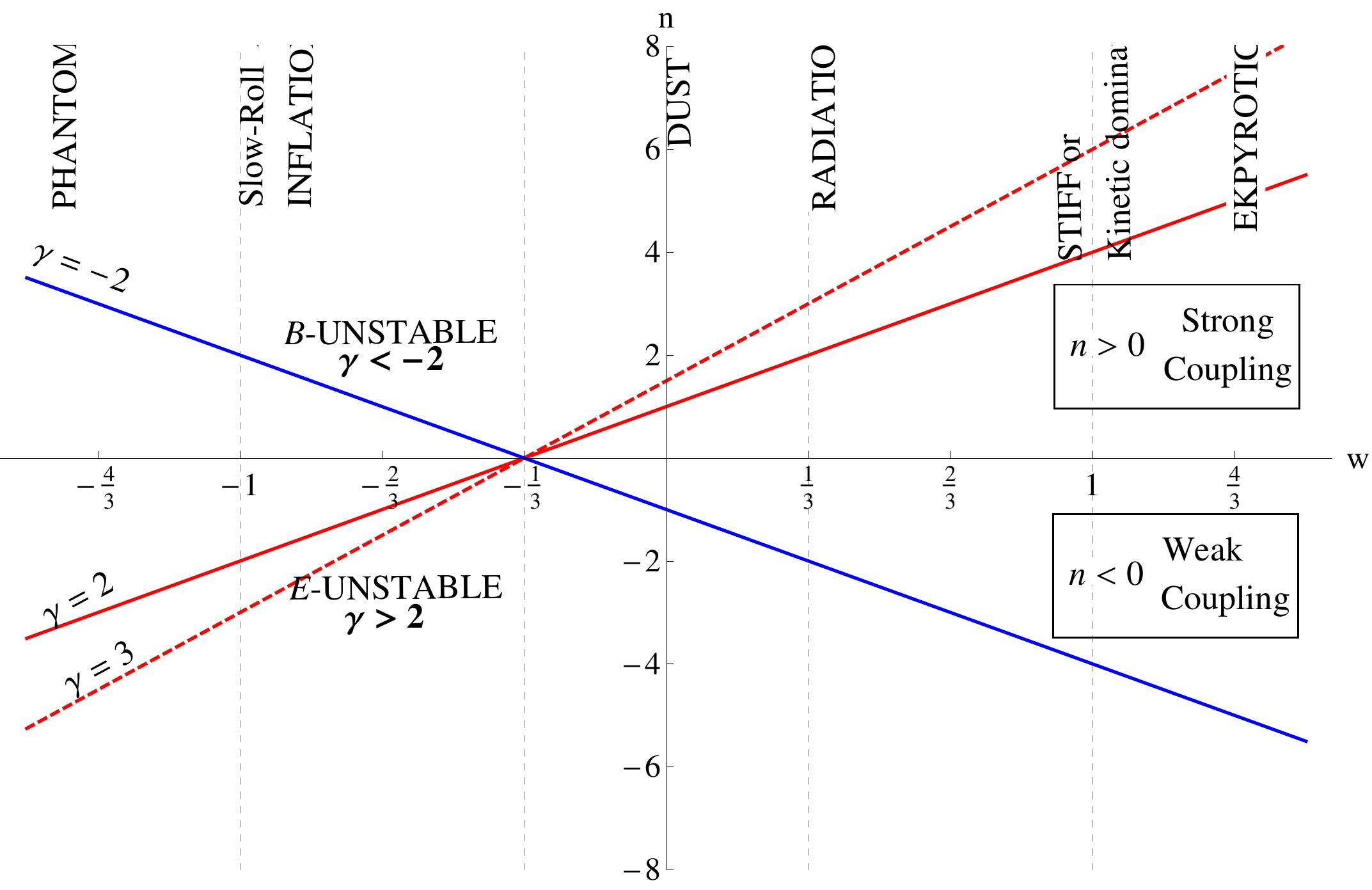}
\caption{The coupling parameter $n$ vs. the equation of state
parameter $w$. The dotted-red line corresponds to $\ga=3$ that
yields scale invariant magnetic fields and red tilted E-spectrum
$n_E=-2$, suffering from strong backreaction. The thick-blue line
is given for $\ga=-2$ that also yields a scale invariant
B-spectrum but with a blue E-spectrum $n_E=2$ with no instability.
The \emph{strong-coupled regime} is defined by the half semiplane
$n>0$, whereas the \emph{weak-coupled regime} belongs to
$n<0$.}\label{nvsw}
\end{figure}

On the contrary, contracting ekpyrotic phases ($w>1$) and bouncing
cosmologies with a contracting phase dominated by pressureless
matter ($w\gtrsim0$), or radiation ($w\simeq1/3$) will not suffer
from this shortcoming, given they belong to the class $w>-1/3$.
Indeed, for this case $\ga$ has the same sign as $n$. Thus, when
we seek for a B-spectrum without the backreaction of the electric
fields, with $\ga<0$, we also obtain $n<0$, that yields the
weak-coupled regime for the theory.

Early contracting periods dominated by an equation of state with
$w>-1/3$ will be potential candidates to succeeded in sustaining
primordial magnetogenesis through vacuum fluctuations of the
kinetically coupled $f^2(\phi)$-electromagnetic theory.

\section{Cosmic magnetic fields through a non-singular bouncing
universe}\label{Bfields}

In the previous section we have seen that it is possible to
overcome the backreaction and strong-coupling problems that are
present in the magnetogenesis models of inflation, by using a
phase of contraction where $w>-1/3$. The weak-coupled regime was
given by the coupling parameter values: $n<0$. This two
assumptions then imply that $\ga<0$, which in turn means that
B-fields dominate the electromagnetic spectrum.

Now we wish to apply the previous result to study the production
of actual magnetic fields originated from an early contracting
universe. In doing so, one has to evolve the super-Hubble
electromagnetic modes through the whole universe history after the
contracting phase.

We shall consider a rather new scenario that starts from a
matter-dominated epoch of contraction \cite{Bran1206}. Here the
primordial matter inhomogeneities spectrum is explained from a
contracting universe in an initial matter dominated state, that
eventually enters in an ekpyrotic contracting phase through a
$\phi$-dominant epoch.  We are particularly interested in such an
example because the dynamics deploy a non-singular bouncing. In
this sense, the evolution of the modes is much simpler than the
singular case. In turn, an important motivation for studying
bouncing cosmologies is to avoid the cosmic singularity problem.

The ekpyrotic phase in contracting cosmologies is of outmost
importance, since it naturally washes away any initial residual
anisotropies that otherwise would grow to destabilize the universe
\cite{ekpyrotic2}.

Another important assumption is that the coupling $f(\phi)$, with
which the electromagnetic fields couple to the background scalar
field that drives the ekpyrotic dynamics, is only relevant during
the ekpyrotic phase, the bounce period and a final kinetic
expanding phase. This means that we would have $f\sim1$ at the
initial matter dominated epoch and at the onset of reheating,
after the kinetic decay of the scalar field. In this sense, though
we are considering a particular model, we shall be testing some
phenomenology about ekpyrotic phases, that appear in many bouncing
cosmologies setups.

However, this may note be the case, and the extension to other
situations where $w>-1/3$, that can somehow control the
anisotropies instability while accounting for observed spectrum of
primordial inhomogeneities $n_s$ and the non-Gaussianity
constraints, seems completely possible and should be further
studied in future works.

\subsection{A background model}

The background model we shall use it was developed in
\cite{Bran1206}, and is an example of the \textit{Matter-Bounce}
scenarios \cite{Bran0904}. It produces a bouncing period, starting
from a contracting ekpyrotic phase to a fast-roll expanding phase.
The dynamics are driven by potential and kinetic energy of a
scalar field $\phi$, by using combined features of Galilean models
and ghost condensate field models. The scalar field Lagrangian
contains higher order derivatives terms of $\phi$, while the
equations of motion remain second order. The Lagrangian density is
 \beg\label{lagrangian}
    \cl_\phi=K(\phi,X)+G(\phi,X)\square\phi,
 \en
where $X\equiv\fr{1}{2}\pa_\mu\phi\pa^\mu\phi$. In this setup the
scalar field is dimensionless. In order to generate the ghost
condensate it is considered that
 \beg\label{Kdephi}
    K(\phi,X)=M_p^2[1-g(\phi)]X+\be X^2-V(\phi),
 \en
the parameter $\be$ guarantees that the kinetic term is bounded
from below at high energy scales, avoiding ghost instability. In
particular, if $g>1$, the bouncing may take place whether
$\dot\phi\neq0$ at that moment. In turn, the value of $g(\phi)$ is
chosen to be negligible when $|\phi|\gg1$ and larger than unity
when $\phi\approx0$. The ansatz used is,
 \beg
    g(\phi)=\fr{2g_0}{e^{-\phi\sqrt{2/p}}+e^{b_g\phi\sqrt{2/p}}}
 \en
where $g_0\equiv g(\phi=0)$ is positive and larger that unity. The
background metric is the usual flat FRWL. Thus, the background
scalar field is spatially homogeneous $\phi=\phi(t)$.

Additionally the Galilean term is fixed so as to stabilize
possible gradient term of cosmological perturbations
\cite{Bran1206}, $G(X)=cX$, with $c>0$.


It is a well known old problem that any contracting cosmological
models have to face the Belinsky-Khalatnikov-Lifshitz (BKL)
chaotic mixmaster behavior \cite{BKL}. The Friedmann equation in
the presence of different types of matter and a curvature term
accounting for flat $\ka=0$, open $\ka=1$ or closed $\ka=-1$,
takes the form
 \beg
    \fr{3}{8\pi}m_{pl}^2H^2=\fr{K}{a^2}+\fr{\rho_m}{a^3}+
    \fr{\rho_{rad}}{a^4}+\fr{\si}{a^6}+\fr{\rho_\phi}{a^{-3(1+w)}}
 \en
where we defined $K\equiv\fr{3}{(8\pi)^2}m_{pl}^4(-3\kappa)$ and
we added a term corresponding to a scalar field.

In an expanding universe the pressureless matter scales slower and
comes to dominate over radiation, also anisotropies are earlier
washed away. The dominant component, however, it would finally be
the curvature term that scales slower than non-relativistic matter
unless the scalar field slow rolls in a flat potential $V(\phi)$,
yielding $w_\phi\simeq-1$. This is the inflationary paradigm to
address the flatness problem.

On the other side, for contracting cosmologies, the anisotropy
energy density increases stronger than matter-radiation and
curvature components. This instability can be mended if we add
another component with an equation of state $w>1$. Thus, the
potential is chosen to yield a contracting ekpyrotic phase during
the downward pull of the scalar field,
 \beg\label{potential}
    V(\phi)=-\fr{V_0}{e^{-\phi\sqrt{2/q}}+e^{b_V\phi\sqrt{2/q}}},
 \en
where $V_0=V(\phi=0)$ has dimensions of (mass)$^4$. This potential
is always negative and is vanishingly small for large values of
$|\phi|$.

The universe starts, in the setup \cite{Bran1206}, from an initial
matter-dominated epoch, with the scalar field on large negative
values far from the potential well $\phi\ll-1$. Also, the
derivative is small $\dot\phi\ll M_p$. However, the potential
makes $\dot\phi>0$ and eventually the scalar field rolls down the
potential. During this period and given that $q<1/3$, one obtains
an ekpyrotic phase where the energy density of $\phi$ scales
faster that anisotropic stresses. Furthermore, to obtain an
ekpyrotic phase, one needs that $\dot\phi\ll M_p$ and $g\approx0$.
In this situation the Lagrangian approaches the usual canonical
form $\cl=M_p^2(\pa_\mu\phi)^2/2-V(\phi)$. The scaling solution,
which is an attractor in phase space for the scalar field, is
given by
 \beg\label{phi ek}
    \phi_{\mathrm{ek}}\simeq-\sqrt{\fr{q}{2}}\ln\lf[\fr{2V_0(t-\tilde t_{b-})^2}{q(1-3q)M_p^2}
    \ri],
 \en
whereas the geometric background dynamics are given by
 \beg\label{a de ek}
    a(t)=a_\star\left(\fr{t-\tilde t_{b-}}{t_{b-}-\tilde t_{b-}}\right)^q,
    \hspace{1cm}H=\fr{q}{t-\tilde t_{b-}}.
 \en
This correspond to mean values of the scale factor and Hubble
parameter during ekpyrosis. The constant $\tilde
t_{b-}=t_{b-}-q/H_{b-}$ is introduced to define a continuous
Hubble parameter from the ekpyrotic contraction to the bounce
phase. In turn, we will have an equation of state from $\phi$,
 \beg\label{EosEkpy}
 w_\mathrm{ek}=-1+\fr{2}{3q},
 \en
As $\phi$ approaches zero, the value of $g$ increases. When
$g(\phi_{b-})=1$ is where the bounce phase starts at time
$t=t_{b-}<0$. During the bounce phase $g(\phi)>1$ and the
quadratic $\dot\phi$ term in (\ref{Kdephi}) becomes negative. At
some point reaches the maximum value $g_0$ at $t=0$ and then
rapidly decreases. At a time $t=t_{b+}>0$ again $g(\phi_{b+})=1$,
the bounce period finishes and the kinetic driven phase starts.
The value of the field at these points may be approximated by
$\phi_{b-}\sim=-\sqrt{p/2}\ln{2g_0}$ and $\phi_{b+}\sim
b_g^{-1}\sqrt{p/2}\ln{2g_0}$, respectively. At the bounce point
the energy density vanishes, which implies the following
approximate relation
 \beg\label{dotphib}
    \dot\phi_b^2\simeq\fr{2M_p^2(g_0-1)}{3\be},
 \en
where it was assumed that $V_0\ll M_p^4$. During the bounce, the
kinetic energy of the scalar field is enhanced to large values,
but during a brief period. Furthermore, as the potential is
bounded from below, in \cite{Bran1206} it was shown that there
cannot arise any possible instabilities at the homogeneous
background level.

After the bounce, $g(\phi)$ decays below unity for $t>t_{b+}$, and
$\dot\phi$ falls back to small values. Since the potential is very
flat for $\phi>\phi_{b+}$, the scalar field will continue
increasing but in a fast-roll phase dominated by kinetic energy,
with an effective equation of state $w_\mathrm{fast}\simeq1$.

The parameters chosen in \cite{Bran1206}, in units of Planck mass,
were:
 \bega\label{param}
\nonumber    V_0=10^{-7},\ \ g_0=1.1, \ \ \be=5,\ \ c=10^{-3},\\
            b_V=5,\ \ b_g=0.5,\ \, p=0.01,\ \ q=0.1
 \ena
with them they found that the maximum amplitude of the Hubble
parameter before (and after) it enters the bounce period is about
$H_{b-}\approx H_{b+}\sim10^{-4}M_p$. Besides, the equation of
state during the ekpyrotic phase is $w\approx 5.67$ for the choice
$q=0.1$.

\subsection{Electromagnetic modes evolution}

\subsubsection{Electromagnetic modes in the ekpyrotic phase: $t<t_{b-}$}

Given that $w_\mathrm{ek}=-1+\fr{2}{3q}$, we obtain for the
parameter (\ref{gamma}),
 \beg\label{gamma ek}
    \ga^{(ek)}=\fr{nq}{1-q},
 \en
regarding that $q<1/3$ solves the anisotropy problem,  we have a
constraint $|n|>2|\ga|$. Apart, the coupling function (\ref{F})
can be expressed as
 \beg\label{F ek}
    \fr{f}{f_\star}=\left(\fr{aH}{a_\star H_\star}\right)^{-\ga}.
 \en
The solution for the electromagnetic modes is
 \beg\label{Sol u Ek}
    u_k^{(ek)}=c_1^{(ek)}(k)+c_2^{(ek)}(k)\left(\fr{a}{a_\star}\right)^{-2n-1+\fr{1}{q}},
 \en
where the corresponding values of $\ga^{(ek)}$ and $w_\mathrm{ek}$
should be used. Furthermore, the E and B-spectrum for modes that
reached the long wavelength regime [see eq. (\ref{drhoB12}) and
(\ref{drhoE})] is given by their spectral indexes
 \beg
    n_B=\left\{%
\begin{array}{ll}
     n_B^{(1)}=4+\fr{2nq}{1-q}, & n<\fr{1-q}{4q} \\
     n_B^{(2)}=6-\fr{2nq}{1-q}, & n>\fr{1-q}{4q} \\
\end{array}%
\right.
   ,\hspace{1cm}n_E=
   \left\{%
\begin{array}{ll}
    n_E^{(1)}=6+\fr{2nq}{1-q}, & n<-\fr{1-q}{4q} \\
    n_E^{(2)}=4-\fr{2nq}{1-q}, & n>-\fr{1-q}{4q} \\
\end{array}%
 \right.\en



The solution for electromagnetic fluctuations generated from
vacuum initial conditions during an ekpyrotic phase is given by
eq. (\ref{Sol u Ek}). This modes have to be matched at the time
$t_{\mathrm{b}-}$ with long wavelength modes in the bouncing
phase. In the same way, this modes will evolve through the
non-singular bounce and match with modes in the fast-roll phase at
the end of the bounce, at $t_{\mathrm{b}+}$.

\subsubsection{A model for $f(\phi)$}

Until now we assumed that the coupling function was time
dependent, and that this time dependence it would be eventually
inherited by the evolution of the scalar field. While the
background evolution is defined through the equation of state $w$,
we expect that the coupling may be defined by a single parameter
$n$ through a power-law. However, when the universe changes phases
(i.e. $w$ changes) it is reasonable that also the coupling
parameter $n$ may change.

To model this time dependence it is then necessary to choose some
specific function $f(\phi)$. Inspired in the models
$e^{\la\phi}F^2$, we consider
 \beg\label{la phi}
    f^2(\phi)=e^{\la(|\phi|-\phi_{\mathrm{end}})}.
 \en
Taking the absolute value of the scalar field amplitude it is a
simplification assumed to keep with the $e^{\la\phi}$-model while
being consistent with the present background dynamics
\cite{Bran1206}. For this case $\phi$ is monotonically increasing
with time, starts from large negative values, goes through zero at
the bounce point, and then reaches a value $\phi_\mathrm{end}>0$.
While, on the other hand, the coupling function $f(\phi)$ (in the
weak-coupled regime) should increase during contraction and
decrease during expansion, reaching its maximum value during the
bounce. In this sense, for the coupling to follow the evolution of
the scalar field properly, a dependence $f=f(|\phi|)$ is needed.
Indeed, with the above definition we obtain $f_{\mathrm{end}}=1$
when $\phi=\pm\phi_{\mathrm{end}}$. Furthermore, when
$\phi=\phi_{\mathrm{b}}=0$ we get the extreme value
$f_{\mathrm{b}}= e^{-\la\phi_{\mathrm{end}}}$.

However, we should remark that such simplification is particularly
problematic at the bounce point, since its time derivative is not
defined there [cf. eq.(\ref{Ec uk})]. Yet, the transition from the
contracting to the expanding branch should be smooth, with
$f(\phi)$ reaching the maximum value at the bounce point, thus
$\dot f=0$. We mend this situation by approximating $|\phi|$ with
a smooth interpolating function, so as to perform the analytic
calculations. We shall consider $|\phi|\sim\phi
\tanh(\phi/(\delta\phi_b))$ where
$\delta\phi_b\ll\phi_{b+}-\phi_{b-}$. Later we verify that the
final result does not depend on this approximation as long as the
bounce is fast enough.

We further remark that the coupling function obtained in previous
work using Weyl integrable spacetimes \cite{Salim0612}, behaves
analytically different from us. In our case we assume a power-law
$f\propto a^n$, thus, in the weak-coupled regime, the coupling
function reaches a maximum at the bounce. They, however, have a
monotonically increasing coupling function, with the time
derivative $\dot f$ reaching a maximum at the bounce. Yet, for
this case there is no power-law spectrum for electromagnetic
fields. Unfortunately their calculations yield a strong-coupled
regime, because their coupling function $I(\omega)=e^{-2\omega}$
remains always well below unity for the parameters $\la\approx
0.07\sim0.1$.

Using the scaling solutions (\ref{phi ek}) and (\ref{a de ek}) we
can express the scale factor as a function of the scalar field,
say $a(\phi)$. By replacing in eq. (\ref{F}) and comparing with
the ansatz (\ref{la phi}), we obtain the relation
 $\sqrt{2}\la=\sqrt{q}n^{(\mathrm{ek})}$. Moreover, it is useful
the generalization to when the phases change to different values
of $q$ (or $w$),
  \beg\label{lamda}
    \sqrt{2}\la=\sqrt{q_1}n^{(1)}=\sqrt{q_2}n^{(2)}=\cdot\cdot\cdot=\mathrm{cte}.
  \en
Then we see that for the weak-coupled regime ($n<0$), we obtain
$\la<0$. Thus, given that $\phi_{\mathrm{end}}>0$, we can check
that $f_\mathrm{b}$ is a maximum, which in turn implies that
$f^{-1}$, related to the coupling 'constant', reaches its minimum
value at the bounce point.

\subsubsection{Modes through the non-singular bounce: $t_{\mathrm{b}-}<t<t_{\mathrm{b}+}$}

As discussed in \cite{Bran1206}, the universe enters the ghost
condensate range at a time $t_{\mathrm{b}-}$. From this time on
the $\dot\phi^2$ term starts to yield a negative contribution that
eventually cancels all the other positive terms in the energy
density. At this moment is when the universe bounces and the
Hubble parameter vanishes.

We need to determine the dynamics of the modes $u_k$ during the
bounce period. The equation for the modes (\ref{Ec uk}) in cosmic
time with the smoothed ansatz for $f^2(\phi)$ yields
 \beg\label{Ec uk bou1}
    \ddot u_k+\lf[ H+\fr{\la\dot\phi}{2}\lf(\fr{\phi/\de\phi}{\cosh(\phi/\de\phi)}+
    \tanh(\phi/\de\phi)\ri)\ri]\dot u_k+\fr{k^2}{a^2}u_k=0,
 \en
clearly we will need an expression for the background quantities,
$H$, $\dot\phi$ and $\phi$ during the bounce. This depends on the
background model we are using. For our case the term proportional
to $\dot\phi$ will show to be negligible for the long wavelength
mode dynamics. We will discuss this point later.

First, it is useful to model the time dependence of the Hubble
parameter during the bounce as
 \beg\label{Hbounce}
    H=\Up t,
 \en
where $\Up$ is a positive constant with $k^2$ dimensions. This
parametrization shows to be valid for a class of fast bounce
models. The bounce time it has been set to zero, $t_b=0$. In turn,
the scale factor is given by
 \beg\label{a de boun}
    a(t)=a_b e^{\fr{\Up t^2}{2}}.
 \en
In this period we do not have an analytical expression for $\phi$,
instead as considered in \cite{Bran1206}, we may use the
analytical approximation for $\dot\phi$,
 \beg\label{dotphi}
    \dot\phi=\dot\phi_b e^{-\fr{t^2}{T^2}}.
 \en
Very close to the bounce, at $t\lesssim0$, one expects that the
modes should reenter the Hubble length for a very brief period of
time. This is because $H$ vanishes at the bounce. Besides, while
$\dot\phi$ reaches it maximum value, the value of
$\phi\rightarrow0$, thus the other term also vanishes. This means
that the modes shift to the short wave regime and back in a very
brief period during the bounce. This same thing happens to the
curvature fluctuations as shown in \cite{Bran1206}. However, this
is not the case at \cite{Salim0612}, because at the bounce
$\dot{f_b}$ is maximum and so the electromagnetic modes remain
bounded to the background dynamics.

\begin{figure}[h]
\centering
\includegraphics[scale=0.8]{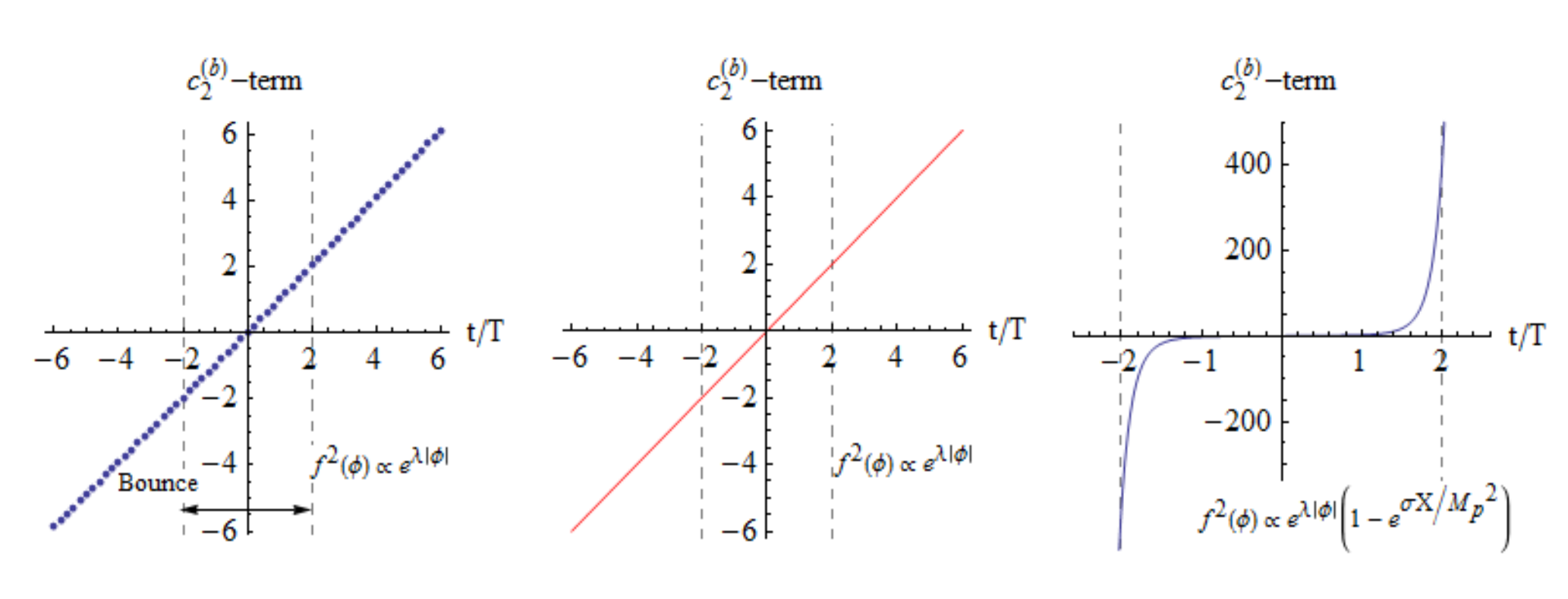}
\caption[Fig]{Left panel: growing modes solved numerically. Center
panel: the linear approximation for $f^2(\phi)\propto
e^{\la|\phi|}$. The plot on the right corresponds to the model
$f^2(\phi,\dot\phi)\propto e^{\la|\phi|}\dot\phi^2/M_p^2$}
\label{Numerical}
\end{figure}

Replacing the approximations (\ref{Hbounce}) and (\ref{dotphi}) in
eq.(\ref{Ec uk bou1}) we obtain for the long wavelength regime
 \beg\label{ec uk bou}
 {u}_{,yy}+\eta(y)
 {u}_{,y}=0,\hspace{1cm}\eta(y)=T^2\Up y+ \la T\dot\phi_b
 e^{-y^2}\fr{|\phi|}{\phi},
 \en
where we shifted to a dimensionless time variable $y=t/T$ and we
abbreviated notation $u_k\equiv u$. The solution comes from the
1st order ODE system
 \beg
    \left\{%
\begin{array}{ll}
    {v}_{,y}+\eta(y){v}=0, \\
    v={u}_{,y}.  \\
\end{array}%
\right.
 \en
One solution is a constant $c^{(\mathrm{b})}_1$. In
\cite{Bran1206} they found $\Up\approx\co(10^{-4})M_p^2$,
$\dot\phi_b\approx0.1M_p$ and $T\approx(t_{b+}-t_{b-})/2$. Using
these, we find that the $e^{-y^2}$ term shows to be negligible as
we go away from $y=0$, and for the long wavelength regime. Thus,
at the onset of the bounce phase and before the kinetic phase
starts, the long wavelengths originated from sub-Hubble
fluctuations during the ekpyrotic phase are determined by
$\eta(y)\approx T^2\Up y$. This was checked numerically for the
present model [see left graphic in figure (\ref{Numerical})]. In
this sense, the time-dependent solution for (\ref{ec uk bou}) is
well described by
$c_2^{(\mathrm{b})}\sqrt{\fr{\pi}{2\la\dot\phi_bT}}
    \mathrm{Erf}\lf[\sqrt{\fr{\la\dot\phi_bT}{2}}y\ri]$. Moreover,
from the figures one can check that, for the brief duration of the
bounce (with $|y|\lesssim2$), it is enough with the linear
approximation $\sqrt{\fr{\pi}{2\la\dot\phi_bT}}
    \mathrm{Erf}\lf[\sqrt{\fr{\la\dot\phi_bT}{2}}y\ri]\approx
    y+\co(\la\dot\phi_b T y^3)$. We can finally write
 \beg\label{Sol u bou}
    u_k^{(\mathrm{b})}=c_1^{(\mathrm{b})}+c_2^{(\mathrm{b})}\fr{t}{T}.
 \en

In the previous analysis we saw that the model had a problem just
at the bounce point that was avoided with a smoothed
approximation. Yet, it may be interesting to consider another
ansatz for the coupling function, that depends with $\phi$ and
$X$. We tried,
 \beg
f^2(\phi,X)=e^{\la(|\phi|-\phi_{\mathrm{end}})}\lf(1-e^{\si
\fr{X}{M_p^2}}\ri),
 \en
with $0<\si<1$. For the present scenario we are considering, the
value of the scalar field vanishes during the bounce $\phi_b=0$.
While on the other hand, the value of its derivative increases as
high as $\dot\phi\approx 0.1M_{p}$. This means that during the
bounce we could approximate the previous expression as
$f^2(\phi,X)\approx
e^{\la(|\phi|-\phi_{\mathrm{end}})}\si\fr{\dot\phi^2}{2M_p^2}$.
Again, by combining this with the approximate analytical
expressions during the bounce of the scale factor(\ref{a de
boun}), the time derivative of the scalar field (\ref{dotphi}) and
the general expression of the coupling function (\ref{F}), we
obtain
 \beg
    \eta(y)=(T^2\Up-4)y+\la T\dot\phi_b
 e^{-y^2}\fr{|\phi|}{\phi}.
 \en
A new $-4y$ term manifests. As we already stated, the last
$e^{-y^2}$ term is negligible in the long wave length regime,
while $T^2\Up\approx T H_{b+}\lf[\fr{1}{2}\ln\lf(\fr{9\be
H^2_{b+}}{(g_0-1)M_p^2}\ri)\ri]^{-1/2}\approx T H_{b+}/2$ for the
set of parameters (\ref{param}) in \cite{Bran1206}. Thus, one
obtains that $\eta(y)\approx-4y$. Which we can easily solve in
terms of the imaginary error function $\mathrm{Erfi}(\sqrt{2}y)$.
This is plotted on the graphic on the right in
fig.(\ref{Numerical}). One can see that during the bounce this
mode is amplified several orders of magnitude. Indeed, for
$-2<y<2$ one obtains an amplification of almost $\co(10^3)$.
Although, we shall leave the analysis of such a possibility for
future work, one can preview that backreaction problems may rise
for a longer bounce period, or if the relative electromagnetic
energy density $\Omega_B$ had grown to levels of about
$\co(10^{-6})$ at the end of the ekpyrotic phase.

\subsubsection{Modes through the fast-roll: $t_{\mathrm{b}+}<t<t_{\mathrm{end}}$}

After the bounce finishes the universe enters a period of
expansion dominated by kinetic energy of the scalar field. In this
stage the equation of state is $w=1$, so it is easy to find the
solution for the electromagnetic modes from the general expression
(\ref{Sol1}).
 \beg\label{Sol u fast}
     u_k^{(\mathrm{f})}=c_1^{(\mathrm{f})}+c_2^{(\mathrm{f})}
     \left(\fr{a}{a_\star}\right)^{r^{(\mathrm{f})}},\hspace{1cm}r^{(\mathrm{f})}=-2n^{(\mathrm{f})}+2
 \en
As before, we have a parameter $\ga^{(\mathrm{f})}$ given by
(\ref{gamma}) using $w=1$, then
 \beg
    \ga^{(\mathrm{f})}=\fr{n^{(\mathrm{f})}}{2}.
 \en
From eq. (\ref{lamda}) we can relate $n^{(\mathrm{f})}$ with
$n^{(\mathrm{ek})}$; by using $q_\mathrm{fast}=1/3$ one obtains
the fast-roll phase equation of state $w_\mathrm{fast}=1$, so
 \beg\label{n fast}
    n^{(\mathrm{f})}=\sqrt{3q_\mathrm{ek}}n^{(ek)}.
 \en
Given that $q_\mathrm{ek}<1/3$ we obtain
$|n^{(\mathrm{f})}|<|n^{(\mathrm{ek})}|$. Thus, as expected, we
will have weak coupling
in the fast-roll expanding phase $n^{\mathrm{(f)}}<0$ and no
backreaction from electric fields $\ga^\mathrm{(f)}<0$.

\subsection{Matching conditions}

In the present nonsingular bouncing model we have two time
surfaces where to match the perturbations originated during the
ekpyrotic phase. The first is before the bounce and after the
ekpyrotic phase, $t=t_{\mathrm{b}-}$, and the second is between
the bounce and the fast-roll phase, $t=t_{\mathrm{b}+}$. The
functions $u_k$ and $\dot u_k$ should be
continuous across the matching surfaces. 
Thus, we have four matching conditions, two at each surface, and
six constants
$c_1^{(\mathrm{ek})},c_2^{(\mathrm{ek})},c_1^{(\mathrm{b})},c_2^{(\mathrm{b})},c_1^{(\mathrm{f})}$
and $c_2^{(\mathrm{f})}$. Furthermore, as $c_1^{(\mathrm{ek})}$
and $c_2^{(\mathrm{ek})}$ are determined by the initial
conditions, the system is completely determined. In fact, to find
the amplitude of electromagnetic fields given by the model, we are
only interested in how the constants
$c_1^{(\mathrm{f})},c_2^{(\mathrm{f})}$ relate to
$c_1^{(\mathrm{ek})},c_2^{(\mathrm{ek})}$. Using the solutions for
ekpyrosis (\ref{Sol u Ek}), the bounce phase (\ref{Sol u bou}) and
the fast-roll period (\ref{Sol u fast}), it is straight forward to
find
 \beg
    \left(%
\begin{array}{c}
  c_1^{(\mathrm{f})} \\
  c_2^{(\mathrm{f})}\\
\end{array}%
\right)=\left(%
\begin{array}{cc}
  m_{11} & m_{12}  \\
  m_{21} & m_{22} \\
\end{array}%
\right)\left(%
\begin{array}{c}
  c_1^{(\mathrm{ek})} \\
  c_2^{(\mathrm{ek})} \\
\end{array}%
\right),
 \en
with
 \beg
   \left(%
\begin{array}{cc}
  m_{11} & m_{12}  \\
  m_{21} & m_{22} \\
\end{array}%
\right)\simeq\left(%
\begin{array}{cc}
  1 & 1+r^{(\mathrm{ek})}H_{\mathrm{b}-}(t_{\mathrm{b}+}-t_{\mathrm{b}-})-\fr{r^{(\mathrm{ek})}H_{\mathrm{b}-}}{r^{(\mathrm{f})}H_{\mathrm{b}+}} \\
  \co(k^2) & \fr{r^{(\mathrm{ek})}H_{\mathrm{b}-}}{r^{(\mathrm{f})}H_{\mathrm{b}+}} \\
\end{array}%
\right),
 \en
where we have used that $t_{\mathrm{b}-}/t_{\mathrm{b}+}\approx
H_{\mathrm{b}+}/H_{\mathrm{b}-}$. Then, we can write the quadratic
amplitude,
 \beg\label{uk 2 fast}
    |u_k^{(\mathrm{f})}|^2\simeq|c_1^{(\mathrm{ek})}|^2+|c_2^{(\mathrm{ek})}|^2
    \lf[m_{12}^2+2m_{12}m_{22}
    \lf(\fr{a}{a_{\mathrm{b}+}} \ri)^{r^{(\mathrm{f})}}+m_{22}^2
    \lf(\fr{a}{a_{\mathrm{b}+}}\ri)^{2r^{(\mathrm{f})}}\ri].
 \en

\subsection{Evolution of seed magnetic fields}

Here we will evolve the amplitude of magnetic fields that reach
the reheating period. For simplicity we are assuming that the
universe reheats instantaneously to the radiation era. Consider
the magnetic cosmological parameter today
 \beg
\Omega^0_B(k)\equiv\fr{\rho_B(t=t_0,k)}{\rho_{\mathrm{cri}}},
 \en
defined at a certain scale $L=2\pi/k$. After the $\phi$ field
disappears the coupling function should stay around
$f_\mathrm{end}\gtrsim1$, the universe enters the radiation
dominated period and the energy density of magnetic fields evolve
as a Maxwell field with $a^{-4}$. Ignoring any type of subsequent
amplification mechanisms until the galaxy formation time, we may
write
 \beg
    \small{\rho_B(t=t_0,k)=\fr{d\rho_B}{d\ln k}(t=t_{\mathrm{end}},k)\left(\fr{a_{\mathrm{end}}}{a_0}\right)^4},
 \en
using the quadratic expression (\ref{uk 2 fast}) in eq.
(\ref{rhoB}) with the constants inherited from the ekpyrotic phase
(\ref{Cte}) we obtain the leading spectral contribution
 \beg
    \Omega^0_B(k)=\fr{4}{3\pi}|b_{(\ga^{(\mathrm{ek})})}|^2\lf(\fr{q}{1-q}\ri)^{2\ga^{(\mathrm{ek})}}
    \fr{H_{\mathrm{end}}^2}{m_{pl}^2}
    \left(\fr{H_0}{H_{\mathrm{end}}}\right)^{2+2\ga^{(\mathrm{ek})}}
    \left(\fr{a_0}{a_{\mathrm{end}}}\right)^{2\ga^{(\mathrm{ek})}}
    \left(\fr{k}{a_0H_0}\right)^{2\ga^{(\mathrm{ek})}+4}.
 \en
To continue we need an expression for $(a_0/a_{\mathrm{end}})$,
and thus, one needs to specify details of the reheating period.
This period is at least described by the initial energy scale
$\rho_{\mathrm{end}}$, the reheating temperature
$T_{\mathrm{reh}}$ and the equation of state $w_{\mathrm{reh}}$.
In fact, all this information can be simplified in a single
parameter $R_{\mathrm{rad}}$ or $R$ described in \cite{Martin07}.
 \beg
     \fr{a_0}{a_{\mathrm{end}}}=\fr{1}{R}\lf(\fr{\rho^0_{\mathrm{rad}}}{M_{p}^4}\ri)^{-\fr{1}{4}}
     \lf(\fr{\rho_{\mathrm{end}}}{M_p^4}\ri)^{\fr{1}{2}}
 \en
where
$\rho_{\mathrm{rad}}^0=\Omega^0_{\mathrm{rad}}\rho^0_{\mathrm{cri}}$.
A complete treatment would involve a study of the parameter space
for different models. However, we shall keep with the simplified
assumption that the reheating is instantaneous. In such a case, we
obtain $R=\rho_{\mathrm{end}}^{1/4}/M_p$, which implies that
$a_0/a_{\mathrm{end}}=(\rho_{\mathrm{end}}/\rho^0_{\mathrm{rad}})^{1/4}$.
Then,

 \beg
    \Omega_B^0(k)=\fr{4}{3\pi}|b_{(\ga^{(\mathrm{ek})})}|^2\lf(\fr{q}{1-q}\ri)^{2\ga^{(\mathrm{ek})}}
    {(\Omega^0_{\mathrm{rad}})}^{-\fr{\ga^{(\mathrm{ek})}}{2}}
    \left(\fr{H_0}{H_{\mathrm{end}}}\right)^{2+\ga^{(\mathrm{ek})}}\fr{H_{\mathrm{end}}^2}{m_{pl}^2}
      \left(\fr{k}{a_0H_0}\right)^{2\ga^{(\mathrm{ek})}+4}.
 \en
Furthermore, the actual magnetic energy density is
$\rho^0_B=B_0^2/2$, then we obtain an amplitude of magnetic fields
 \beg
    \fr{B_0}{m_{pl}^2}=\fr{|b_{(\ga^{(\mathrm{ek})})}|}{\pi}\lf(\fr{q}{1-q}\ri)^{\ga^{(\mathrm{ek})}}
    {(\Omega^0_{\mathrm{rad}})}^{-\fr{\ga^{(\mathrm{ek})}}{4}}
    \left(\fr{H_0}{m_{pl}}\right)^{1+\fr{\ga^{(\mathrm{ek})}}{2}}
    \lf(\fr{H_{\mathrm{b}+}}{m_{pl}}\ri)^{-\fr{\ga^{(\mathrm{ek})}}{2}}e^{-\fr{3\ga^{(\mathrm{ek})}N_{\mathrm{f}}}{4}}
    \left(\fr{k}{a_0H_0}\right)^{\ga^{(\mathrm{ek})}+2}
 \en
where we used that
$H_{\mathrm{end}}=H_{\mathrm{b}+}e^{\fr{3}{2}N_{\mathrm{f}}}$. We
may express the magnetic fields in \emph{gauss} through
$1m_{pl}^2=7.64\times10^{57}\mathrm{G}$. The measured actual
Hubble parameter is close to
$H_0=70\mathrm{km}/s\mathrm{Mpc}^{-1}=1.23\times10^{-61}m_{pl}$\cite{Planck13cosmoparam},
and the radiation ratio
$\Omega^0_{\mathrm{rad}}=3.53\times10^{-5}$
\cite{reviewparticle12}. Considering these, we arrive at

 \bega\label{B0}
\nonumber
\left(\fr{B_0}{\mathrm{gauss}}\right)\simeq\fr{7.64\times10^{57}
    (1.23\times10^{-61})^{2+\fr{\ga^{(\mathrm{ek})}}{2}}}{(3.53\times10^{-5})^{\fr{\ga^{(\mathrm{ek})}}{4}}}
    &&\fr{|b_{(\ga^{(\mathrm{ek})})}|}{\pi}\lf(\fr{q}{1-q}\ri)^{\ga^{(\mathrm{ek})}}e^{-\fr{3\ga^{(\mathrm{ek})}N_{\mathrm{f}}}{4}}\\
    && \lf(\fr{H_{\mathrm{b}+}}{m_{pl}}\ri)^{-\fr{\ga^{(\mathrm{ek})}}{2}}
     \left(\fr{2.7\times10^{4}\mathrm{Mpc}}{L}\right)^{\ga^{(\mathrm{ek})}+2},
 \ena
an expression for actual magnetic fields at scales $L$ measured in
\emph{Mpc} and produced during an ekpyrotic contracting phase with
equation of state $w_\mathrm{ek}=-1+2/3q>1$. In deriving it, we
have made several assumptions. These assumptions involve the
particular background evolution of the scalar field that drive the
ekpyrotic phase, the bouncing, and the final fast-roll phase. We
also considered an instantaneous reheating and, furthermore, we
supposed that the magnetic fields are not amplified (nor consumed)
by any other mechanism during the universe evolution until galaxy
formation.

\begin{figure}[h]
\centering
\includegraphics[scale=1.]{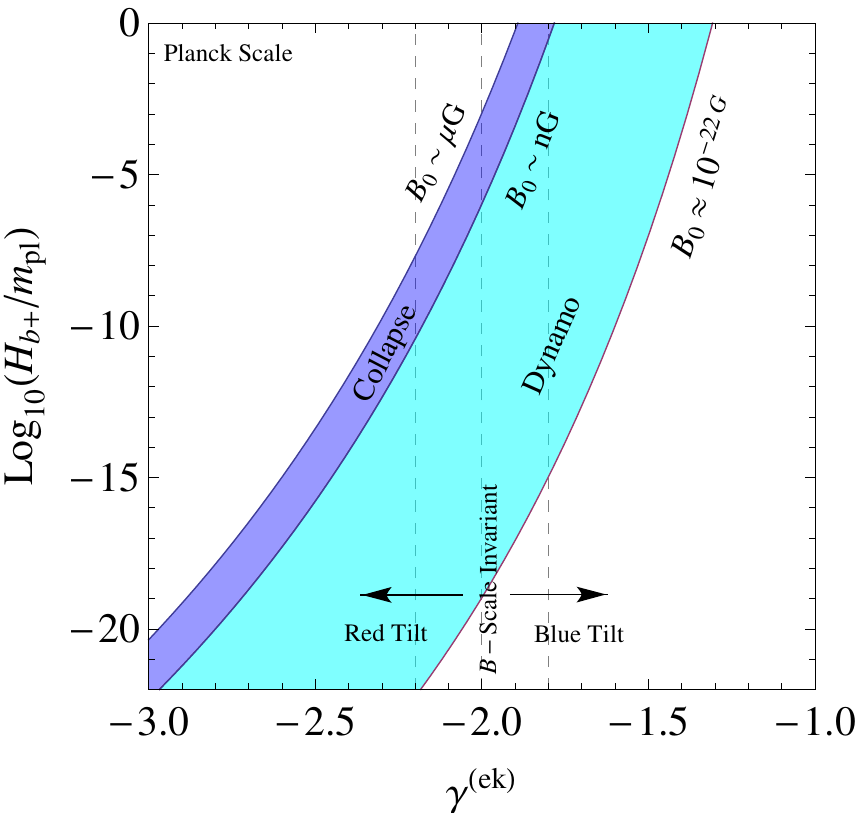}
\caption[Fig]{$\mathrm{Log}H_{b+}$ vs. $\ga^\mathrm{(ek)}$ from
expression (\ref{B0}) for $B_0$. We included the magnetic fields
produced at $L=1Mpc$. The blue belt corresponds to the interval
between $B_0=10^{-6}G$ and $10^{-9}G$, that account for
compression of B-field lines by the collapse of the protogalaxy.
The cyan belt corresponds to the dynamo mechanism needed levels up
to $10^{-22}G$.} \label{fig.B}
\end{figure}

The B-spectrum it is given by the value of $\ga^{(\mathrm{ek})}$
at the ekpyrotic phase. Its amplitude depends also on the values
of the equation of state during that period through the value of
$q$ and of the energy scale $H_\mathrm{end}$ at the onset of the
reheating period. In Fig.\ref{fig.B} we use this expression
(\ref{B0}) to identify the regions where interesting magnetic
fields maybe produced by this mechanism. We included magnetic
fields between microgauss, corresponding  to the actual
observations on galactic scales and nanogauss, the last needed if
amplification is only done by the collapsing protogalaxy. Besides,
nanogauss fields correspond to upper limits inferred by Planck
constraints on non-Gaussianity \cite{Plancknongauss} from the
bispectrum \cite{bispectrum} and recently from the trispectrum
\cite{trispectrum}. The cyan belt corresponds to the dynamo needs
for amplifying primordial magnetic fields of up to $10^{-16}\mu$G
to $\mu G$ levels.

\begin{figure}[h]
\centering
\includegraphics[scale=0.8]{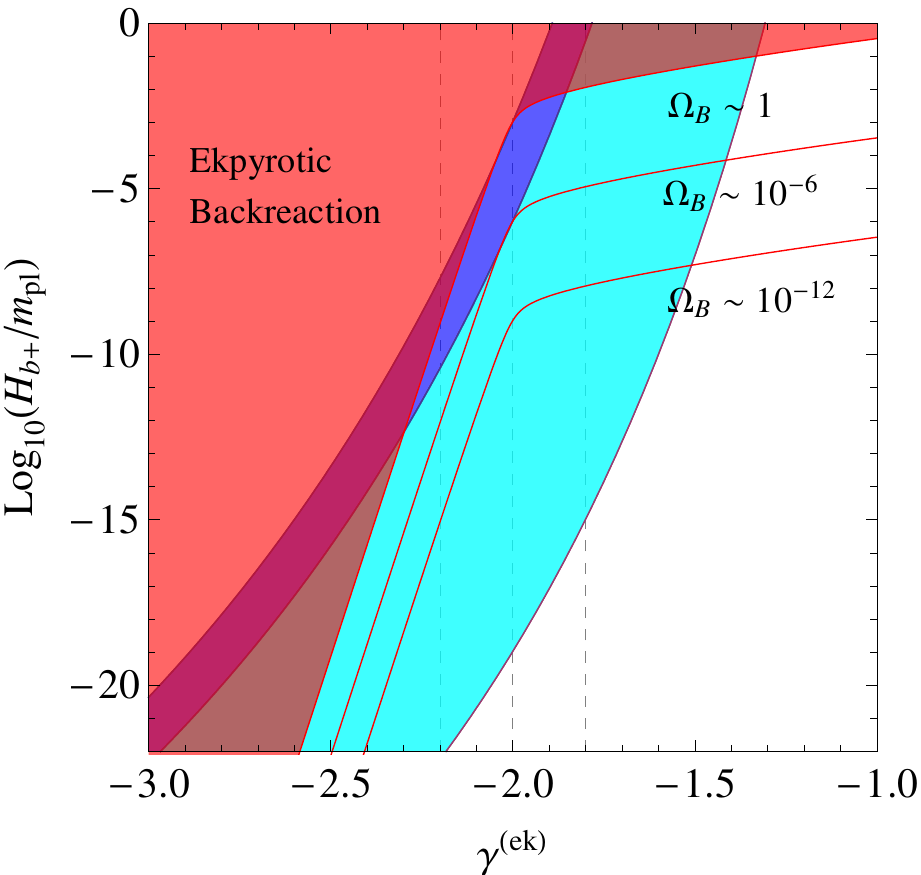}
\caption[Fig]{ $\mathrm{Log}H_{b+}$ vs. $\ga^\mathrm{(ek)}$, with
backreaction levels during ekpyrosis of $N_\mathrm{ek}=75$ by
using eq.(\ref{OmegaB1}). The red region is forbidden since there
$\Omega_B>1$. Red lines correspond to allowed levels of
backreaction $\Omega_B=10^{-6}$ and $\Omega_B=10^{-12}$. We also
included the magnetic fields produced at $\sim1Mpc$ as in fig.
(\ref{fig.B}). Here, we considered that fast-roll is very short
$N_\mathrm{f}\approx0$ and $q=0.1$.} \label{back1}
\end{figure}

On Fig.\ref{back1} we included the backreaction constraints from a
period of contraction using Eq.(\ref{OmegaB1}). Though we used the
value $q=0.1$, that corresponds to an ekpyrotic phase with
$w\approx5.67$, one may go for values of radiation or
matter-dominated contractions without significant variations.
Also, in the present model we have
$H_{b+}\approx10^{-4}m_\mathrm{pl}$, thus one obtains strong
magnetic fields $\gtrsim 10^{-3}\mu$G if the B-spectrum is almost
scale invariant $n_B\approx0$. But, also for
$B_\mathrm{1Mpc}$-fields with blue tilt up to $n_B=1$, one can get
strengths $\sim10^{-16}\mu$G at safe levels of backreaction
$\Omega_B\approx10^{-6}$.

Another interesting feature is that apparently one can relax the
scale of the bouncing up to about
$H_{b+}\approx10^{-15}\sim10^{-20}m_\mathrm{pl}$, keeping a low
blue tilt $n_B=0\sim0.4$, and still obtain interesting magnetic
fields.

We note, however, that in these last calculations we ignored a
durable fast-roll phase given that we used $N_{\mathrm{f}}=0$.

\subsection{Red-tilt and B-instability}

Initially the background model should cast for the solution of the
cosmological problems of the Big-Bang. In particular, the flatness
problem is solved if $aH$ grows at least $60$ e-folds in
inflationary setups. However, in general, bouncing cosmologies
address the flatness problem naturally, given that during the
contracting phase curvature relative energy $\Omega_k$ decays with
respect to almost every component in the universe (except, of
course, a cosmological constant). Indeed, we checked that if the
ekpyrotic phase is rather short, $N_\mathrm{ek}\lesssim30$, the
constraints on the red-tilted spectrum are significatively
relaxed.

\begin{figure}[h]
\centering
\includegraphics[scale=0.75]{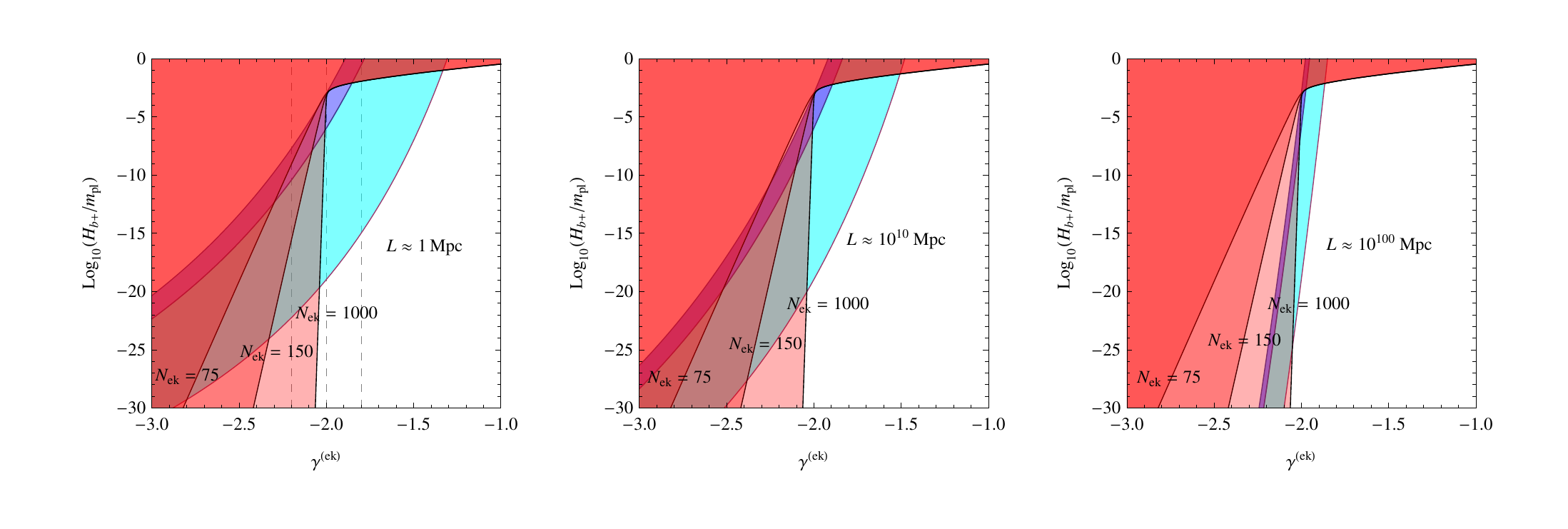}
\caption[Fig]{Regions of $\mathrm{Log}H_{b+}$ vs.
$\ga^\mathrm{(ek)}$ for backreaction ($\Omega_B\geqslant1$) during
different lasting periods of ekpyrosis, $N_\mathrm{ek}=75, 150,
1000$ (red shaded regions), and for actual magnetic fields $B_0$
with lengths of coherence, from left to right, $L=1, 10^{10}
\mathrm{and}10^{100}$Mpc.} \label{variosN}
\end{figure}

To proceed, we would like to test how the system is sensible to
develop instabilities for larger values of $N_{\mathrm{ek}}$. In
Fig.\ref{variosN} we sketched how backreaction grows for different
values of $N_{\mathrm{ek}}$. We identify that the effect is rather
weak, and are needed large numbers of e-folds,
$N_{\mathrm{ek}}\approx1000$ to practically exclude the red tilted
B-spectrum of magnetic fields. While the blue tilted region
remains totally unharmed.  We also showed the qualitative behavior
of the amplitude of magnetic fields in the strengths of interest,
for collapse of the protogalactic cloud: $\mu$G to $n$G (blue
belt) and dynamo mechanism: $n$G to $10^{-22}$G (cyan belt), when
we vary their scales of coherence $L$. We included extremely large
values, in the second and third panels ($L=10^{10}\mathrm{Mpc}$
and $L=10^{100}\mathrm{Mpc}$, senseless otherwise one assumes
longlasting periods $N_{ek}\gg100$) only to illustrate the
weakness of the effect. Clearly, we will be interested in the
usual values about $N_{\mathrm{ek}}=75$, with B-fields at a
coherence scale of $1Mpc$. Yet, variations around these values
will not change significantly the general behavior.

However, we have not considered the possible instabilities during
the kinetically-driven expanding phase. Nor consider how B-fields
decay during this period. These will be address in the next
section.

\subsection{Backreaction during the fast-roll}

The calculation is very similar to what we have done previously.
First we need to determine the energy densities (\ref{rhoB}) and
(\ref{rhoE}) using $u_k^{(\mathrm{f})}$. As seen before, the
constants of the mode $u_k^{(\mathrm{f})}$ are related, through
the matching conditions, to the constants of the ekpyrotic mode
$u_k^{(\mathrm{ek})}$. These quantities were fixed by initial
conditions during the ekpyrotic contracting phase, and are given
by eq. (\ref{Cte}) with the ekpyrotic values $\ga^{(\mathrm{ek})}$
and $w=-1+2/(3q)$.

Additionally, to calculate the backreaction during this period we
need to consider the interval of modes that had exited the Hubble
length during the ekpyrotic contraction, and start to reenter
during the fast roll phase. The initial interval to consider, it
will then be given approximately by the same interval that
suffered the transition from the sub-Hubble regime to the
super-Hubble regime. We characterized this interval through the
effective number of e-folds
$N_{\mathrm{ek}}=\ln\lf(\fr{a_{\mathrm{b}-}H_{\mathrm{b}-}}{a_{ek}H_{ek}}\ri)$.
Afterwards, as the fast-rolls period develops, comoving
wavelengths reenter as the comoving Hubble length $(aH)^{-1}$
increases. This means that the effective number of e-folds
$N_{\mathrm{f}}=\ln\lf(\fr{a_{\mathrm{end}}H_{\mathrm{end}}}{a_{\mathrm{b}+}H_{\mathrm{b}+}}\ri)$
will be negative during this phase. Clearly, if the fast-roll
period is almost instantaneous, then $N_\mathrm{f}\approx0$ and
there are no modes reentering the Hubble sphere that can backreact
on the background.

We will just keep the dominant contributions that generates
instabilities during this period. We find that these are produced
by the last $c_2^{(\mathrm{ek})}$-term proportional to $m_{22}^2$
in (\ref{uk 2 fast}). The reason of this is not so difficult to
understand given that during this phase the scale $a$ grows and
$r^{(\mathrm{f})}=2-2n^{(\mathrm{f})}>0$. Furthermore, it is
straightforward to check that electric fields also have an
instability from the same term. Moreover, these two contributions
are proportional and they e-fold with the same rate. Thus, we find
that
 \bega\label{Omegafast}
\nonumber
\Omega_{B}^{(\mathrm{fast})}+\Omega_E^{(\mathrm{fast})}&\simeq&\fr{4}{3\pi}m_{22}^2|b_2(\ga^{(\mathrm{ek})})|^2
    \lf(\fr{1}{6-2\ga^{(\mathrm{ek})}}+\fr{4r^{(\mathrm{f})2}}{4-2\ga^{(\mathrm{ek})}}\ri)
    \lf(\fr{q}{1-q}\ri)^{2-2\ga^{(\mathrm{ek})}}
    \times\\
    &&\fr{H^2_{\mathrm{b}+}}{m_{pl}^2}e^{[3+2(\ga^{(\mathrm{f})}-
    \ga^{(\mathrm{ek})})]N_{\mathrm{f}}}.
 \ena
These are the only contributions that can grow during the
fast-roll. The others just decay with different rates and are
negligible.

It is found that $\ga^{(\mathrm{f})}<\ga^{(\mathrm{ek})}<0$, and
noticing that for $q_\mathrm{ek}=1/3$ one gets
$\ga^{(\mathrm{ek})}=\ga^{(\mathrm{f})}$, then the exponential
rate is restricted to
$3+2(\ga^{(\mathrm{f})}-\ga^{(\mathrm{ek})})<3$. We see that still
there may be stability for this components as long as this rate
keeps positive (one should remind that $N_\mathrm{f}<0$). This
depends strongly on the parameter $q$ of the ekpyrotic phase, and
more stability is obtained as close as $q=1/3$ we are. In another
way, for sufficiently steep ekpyrotic potentials, strong
instabilities may arise in a durable kinetic stiff dominated
expanding phase of the scalar field. Thus, care should be taken
for models where $q\ll1$, because they are susceptible to develop
$f^2(\phi)F^2$- instabilities.

\begin{figure}[t]
\centering
\includegraphics[scale=.75]{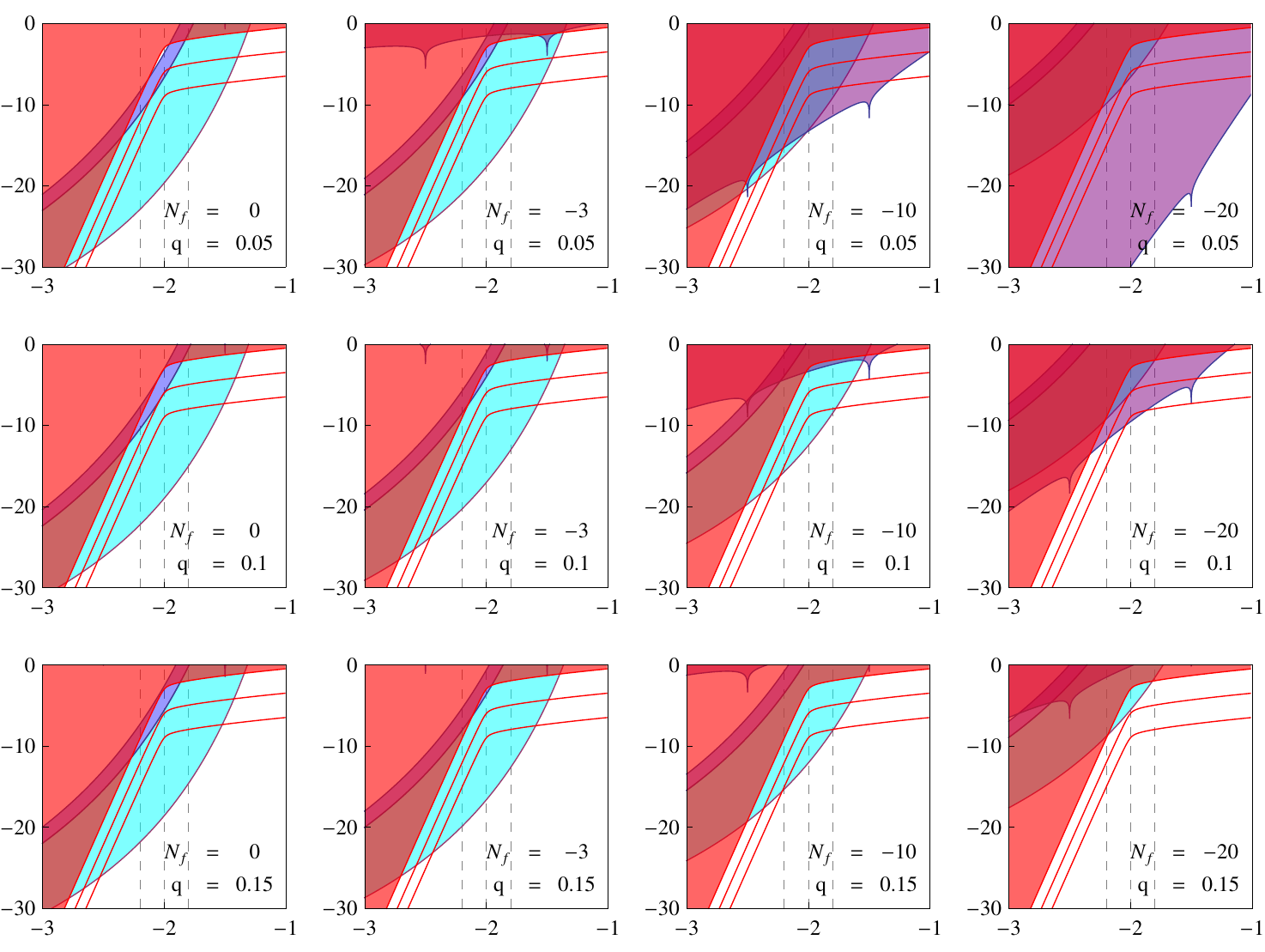}
\caption[Fig]{The scale of the bounce
$Log_{10}(H_{\mathrm{b}+}/m_{pl})$ vs. the parameter
$\ga^{(\mathrm{ek})}$. In rows, from top to bottom, we listed
$q=0.05,0.1$ and $0.15$. In columns, we have from left to
right,the duration of the kinetic period $N_{f}=0,-3,-10$ and
$-20$. The red sector corresponds to
$\Omega^{(\mathrm{ek})}_{EM}>1$, backreaction produced during a
$N_\mathrm{ek}=75$ ekpyrotic contraction phase. The purple sector
corresponds to $\Omega^{(\mathrm{fast})}_{EM}>1$, backreaction
during the fast-roll expansion. We included the magnetic fields
produced at $\sim1Mpc$. The blue belt corresponds to
$B\sim10^{-6}G$ to $10^{-9}G$.} \label{back2}
\end{figure}

\subsection{Magnetic fields without backreaction}

In previous sections we quantified the backreaction levels of
magnetogenesis that operate during an ekpyrotic phase [see
eq.(\ref{OmegaB1})] and during the fast-roll phase [see
eq.(\ref{Omegafast})]. We argued before that backreaction during
the contracting phase is not so sensible to the value of $q$, as
long as $w>-1/3$. However, this is not the case for the expanding
fast-roll phase. As one may check in Fig.(\ref{back2}), the purple
region corresponds to forbidden regions of backreaction that
originate during the fast-roll phase. The values used in the
present model correspond to the middle row, where $q=0.1$. In this
case backreaction excludes any magnetic fields whether $(aH)$
e-folds almost $N_\mathrm{f}=-20$. Instead, for $N_\mathrm{f}=-10$
still a large window of parameters allows magnetogenesis. In
particular, for $N_\mathrm{f}=-3$ and almost scale invariant
B-fields one obtains nanogauss strengths for
$H_{b+}\approx10^{-4}\sim10^{-5}$. Yet, the situation changes
drastically when we go for lower values of $q$. For example, for
$q=0.05$ (first row in fig.(\ref{back2})) or $w\simeq12.3$, we
find that for $N_\mathrm{f}=-10$, backreaction has practically
excluded all possible parameters. In contrast, when $q=0.15$ (last
row) which corresponds with $w\simeq3.4$, it is safe from
backreaction. This behavior is related to the differences between
the equations of state during the contracting and expanding
phases. In this sense, we find that backreaction is enhanced when
this differences are stronger. In our case, when $q$ becomes much
smaller that $1/3$. From the potential point of view, a lower
value of $q$ corresponds to steeper potentials, while during the
fast-roll phase it is assumed that the potential is much more
flat. In turn, this would mean that the system is unstable against
strong differences in the slopes of the potential between the
downward and the upward branch.\\

In this sense we find that in scenarios with an ekpyrotic
contraction and a fast-roll-stiff period, magnetogenesis is
possible as long as $q\lesssim1/3$. It is remarkably that in
\cite{Bran1206} they found that the spectral index of thermally
seeded  scalar fluctuations is $n_s=4q/(1-q)$. Thus, one obtains
the slightly red spectrum for very close values below $q=1/5$. In
this case stability is guaranteed.

In a more general framework we may think that instabilities appear
related to the asymmetries of the potential $V(\phi)$. Or in
another way, symmetric potentials are stable.

\section{Conclusions}\label{Conclu}

In the present work we have analyzed the primordial magnetogenesis
issue from the coupling of a scalar field to a $U(1)$ gauge field.
Particularly we focused our study in the $f^2(\phi)F^2$-class
models.

The first part of the paper was devoted to analyze the known
problems of backreaction and strong coupling that plague this
scenarios in the inflationary case. As a result we found that, in
general, inflation cannot support magnetogenesis without going to
the strong coupling regime. This was realized since we found that
when $w<-1/3$ (required by any inflationary model), one obtains
different signs of $\ga$ and $n$. Indeed, we showed that for the
values $\ga<0$, B-fields dominate the spectrum, thus we speak of
magnetogenesis. On the other side, $\ga>0$ makes E-fields lead the
spectrum and electrogenesis is achieved. Furthermore, the strong
coupled regime is for $n>0$. In this sense, in inflationary
scenarios one should keep $n<0$ which automatically yields
$\ga>0$. This means looking for B-fields in a place where E-fields
dominate and strong instabilities from backreaction arise.

On the other side, both problems can be solved whether we have an
early cosmological period with $w>-1/3$ (meaning $\ga$ and $n$ of
the same sign), and with the coupling $f(\phi)$. Such candidates
belong to bouncing cosmologies, that have a period of contraction,
which in order to generate large scale fluctuations, need to have
a shrinking Hubble sphere during contraction. Contrary to an
expanding inflation cosmology, this is achieved when $w>-1/3$.

Consequently, the second part of our work consisted in applying
this arguments to a specific bouncing cosmology scenario. We chose
a model where the bouncing is non-singular \cite{Bran1206}, that
makes it much easier to follow the modes through the different
phases. Here, the primordial spectrum of matter inhomogeneities is
addressed by considering an initial matter-dominated phase
($w\simeq0$) that evolves through an ekpyrotic period of
contraction that washes out anisotropies present at the start
\cite{Bran1301}. All the work is done by a background scalar field
in an ekpyrotic-like potential. After, when the scalar field gets
close to the minimum of the potential, a brief ghost condensate
phase turns on, during which the universe bounces. Finally, the
scalar field enters a flatter branch of the potential, dominated
by kinetic energy, yielding a stiff-like period of expansion,
where the scalar field decays with respect to the other components
of the universe regarding the initial conditions for reheating and
the standard Big-Bang.

What we done was to introduce the $f^{2}(\phi)F^2$ theory in the
background bouncing dynamics. Though the scenario starts from an
initial matter-dominated phase, we decided to describe only the
electromagnetic fields generated by vacuum fluctuations during the
ekpyrotic phase. The reason for this was to keep as simple and
general as possible, given that ekpyrotic contracting periods
before the bounce are commonly needed conditions to evade the
instabilities of anisotropies, and belong to many bouncing
universe models. However, we suppose that the extension to
consider other previous phases to ekpyrosis should not in general
be so problematic as long as $-1/3<w<1$. Next we needed to choose
a specific coupling function $f(\phi)$ of the $e^{\la\phi}$ type
[see eq. (\ref{la phi})]. With it we obtained the relation
(\ref{lamda}) between the coupling parameter $n$ and the equation
of state $w=-1+2/(3q)$. Finally, we could define the solutions for
the modes in the long wavelength regime for ekpyrosis, bounce, and
fast-roll phases.\\
At the same time we computed the relative energy density of the
electromagnetic field during ekpyrosis and the fast-roll expanding
phase.

This was necessary to follow the levels of backreaction of the
perturbations during its evolution. On the other side,
instabilities during the bounce period are negligible. This should
be clear from eq. (\ref{Sol u bou}), where the solution for the
modes is practically linear and the duration of the bounce is very
short. However, this could turn to be a very particular case.
Indeed, as we showed in another example, one can obtain large
amplification even during short bouncing phases.

When analyzing backreaction we first found that the duration of
the ekpyrotic contracting phase, characterized in the effective
number of e-folds of $a|H|$, can at least exclude the red tilted
spectrum $(\ga^\mathrm{(ek)}<-2)$ if it goes for more than about
$\sim1000$ e-folds. But it left the blue tilted region of the
parameters unconstrained [see fig.\ref{variosN}]. Indeed, such a
behavior seems compatible with cyclic scenarios \cite{cyclic}.

Moreover, a newly interesting feature is that the fast-roll
expanding period could develop strong instabilities of
backreaction if $q\ll1/3$. This would mean that strong ekpyrotic
periods could turn to be unstable for such theories. Indeed, a
related study involving $p-$forms was previously done in
\cite{Steinhardt0312}. On the other side, as in the present
bouncing model, weaker ekpyrotic phases favor a safer
magnetogenesis.


From a physical point of view, a stronger ekpyrotic phase can be
identified with a steeper potential for the scalar field. In this
sense, as we noticed when considering the backreaction during the
stiff fast-roll phase, instabilities appear for steeper
potentials. This means that stability is obtained for rather
symmetric potentials.

Finally, to complete our work we estimated present amplitude of
magnetic fields generated from this early epochs. As expected a
broad range of parameters can account for the needed strengths of
B-fields. This should be identified with the fact that contracting
cosmologies allowed us to work with the magnetogenesis branch
$\ga<0$. As was previously observed, for this case B-fields
dominate the spectrum of the electromagnetic fields.

Finally we expect that contracting universes dominated by matter
or radiation, previous to the ekpyrotic washing of anisotropies,
can also account for amplification of vacuum fluctuations of the
gauge field (as long as the coupling $f(\phi)$ exists). Indeed,
given that the Hubble sphere shrinks when $\dot a<0$ and $w>-1/3$,
an interval of wavelengths suffers the 'freezing' on cosmological
scales, go through the ekpyrotic phase, the bounce and reenter on
late times. In such a case, one immediately sees [cf.
eq.(\ref{lamda})] that the coupling parameter remains in the weak
coupling regime through all the evolution. Apart, a bouncing phase
can also originate from pure quantum cosmological effects. They
also provide a natural framework in which a dust-dominated
contraction is easily implemented to yield a scale-invariant
spectrum of perturbation \cite{PintoNeto}.

The recent BICEP2 experiment data release \cite{BICEP2} detected
cosmological gravitational waves over $5\si$, from measurement of
the B-mode polarization of the CMB. The tensor-to-scalar ratio
found is $r=0.2^{+0.07}_{-0.05}$. In this sense, the original
ekpyrotic and cyclic universe scenarios, that predict very low
gravitational wave production, are strongly constrained.

However, we would like to remark that in the present paper we
find, as a general result, that early contracting cosmological
phases with $w>-1/3$ support magnetic field generation safe from
the backreaction/strong-coupling problem. Indeed, it has been
argued that combining BICEP2 and POLARBEAR \cite{POLARBEAR}
polarization data, a preference for blue tensor spectrum is
obtained \cite{Xue1403}. In particular, we considered a bouncing
universe example that belongs to the matter-bounce scenarios. In
this realization, tensor perturbations are nearly scale-invariant
but the amplitudes too high. Yet, from the two-field picture
\cite{Bran1305} the tensor modes become a tunable parameter. Thus,
it should be further studied if this background model fits the
observations. Moreover, we expect that bouncing models that can
fit the observed tensor-to-scalar ratio will also be able to
support magnetogenesis.

\section*{Acknowledgments}

I wish to thank Yi-Fu Cai, Robert Brandenberger, Alexander Kusenko
and Anupam Mazumdar for useful comments and clarifications. Also I
am very grateful with Jose M. Salim for useful discussions and
hospitality at the CBPF-Rio de Janeiro. This work was done at the
CBPF-Rio de Janeiro supported by 2012 CNPq-TWAS Postdoctoral
Fellowship Programme, FR number: 3240226623.

\end{document}